\documentclass[10pt,pre,aps,twocolumn,showpacs,superscriptaddress,
floatfix]{revtex4}

\usepackage{graphicx}

\newcommand{\be}{\begin{equation}}
\newcommand{\ee}{\end{equation}}
\newcommand{\bea}{\begin{eqnarray}}
\newcommand{\eea}{\end{eqnarray}}
\newcommand{\down}{\downarrow}
\newcommand{\up}{\uparrow}

\newcommand{\f}{\frac}

\newcommand{\Deltatau}{\Delta \tau}

\begin{document}

\title{Quantum Monte Carlo with Directed Loops}

\author{Olav F. Sylju{\aa}sen}
\email{sylju@nordita.dk}
\affiliation{NORDITA, Blegdamsvej 17, DK-2100 Copenhagen {\O}, Denmark}

\author{Anders W. Sandvik}
\email{asandvik@abo.fi}
\affiliation{Department of Physics, {\AA}bo Akademi University, 
Porthansgatan 3, FIN-20500 Turku, Finland}

\date{\today}


\preprint{NORDITA-2002-13CM}

\begin{abstract}
We introduce the concept of directed loops in stochastic series expansion and 
path integral quantum Monte Carlo methods. Using the detailed balance rules 
for directed loops, we show that it is possible to smoothly connect generally 
applicable simulation schemes (in which it is necessary to include 
back-tracking processes in the loop construction) to more restricted loop 
algorithms that can be constructed only for a limited range of Hamiltonians
(where back-tracking can be avoided). The ``algorithmic discontinuities'' 
between general and special points (or regions) in parameter space can hence 
be eliminated. As a specific example, we consider the anisotropic $S=1/2$
Heisenberg antiferromagnet in an external magnetic field. We show that 
directed loop simulations are very efficient for the full range of magnetic 
fields (zero to the saturation point) and anisotropies. In particular for 
weak fields and anisotropies, the autocorrelations are significantly reduced 
relative to those of previous approaches. The back-tracking probability 
vanishes continuously as the isotropic Heisenberg point is approached. For 
the XY-model, we show that back-tracking can be avoided for all fields 
extending up to the saturation field. The method is hence particularly 
efficient in this case. We use directed loop simulations to study the 
magnetization process in the 2D Heisenberg model at very low temperatures. 
For $L\times L$ lattices with $L$ up to $64$, we utilize the step-structure
in the magnetization curve to extract gaps between different spin sectors.
Finite-size scaling of the gaps gives an accurate estimate of the 
transverse susceptibility in the thermodynamic limit: 
$\chi_\perp = 0.0659 \pm 0.0002$.
\end{abstract}

\maketitle


\section{introduction}
In recent years, significant advances in quantum Monte Carlo (QMC) algorithms 
have opened up several classes of quantum many-body models to the kind of
large-scale numerical studies that were previously possible only for classical
systems. The progress has been along two main lines: (i) the elimination 
\cite{sse1,sse2,prokofev,beard,rombouts} of the systematic error of the 
Trotter decomposition \cite{suzuki1} on which most of the early 
finite-temperature QMC algorithms 
\cite{suzuki2,barma,worldline,cullen,determinant} were based (with the
exception of Handscomb's method \cite{handscomb,lyklema,lee,chakravarty}, 
the utility of which was limited), and (ii) the development of loop-cluster 
algorithms \cite{evertz} for efficient sampling in the quantum mechanical 
configuration space \cite{prokofev,beard,kawashima,sse3,syljuasen}. 
Algorithms incorporating both (i) and (ii) have been devised starting from 
either the Euclidean path integral (world-line QMC methods operating in 
continuous imaginary time \cite{prokofev,beard}) or the power series expansion
of the partition function (stochastic series expansion, hereafter SSE 
\cite{sse3}, which is an extension of Handscomb's method). While the Trotter 
error is a controllable one and can be eliminated in standard approaches by 
extrapolating results for different imaginary time discretizations to the 
continuum \cite{suzuki1,fye}, its absence directly at the level of the 
simulation can imply considerable time savings when unbiased 
results are needed, e.g., in finite-size scaling studies. The
loop-cluster algorithms (world-line loops \cite{evertz,kawashima,syljuasen}, 
SSE operator-loops \cite{sse3}, and world-line worms \cite{prokofev}) have 
offered even more dramatic speed-ups, in many cases reducing autocorrelation 
times by several orders of magnitude and thus enabling studies of system sizes
much larger than what was possible with local sampling algorithms. 
In addition, in some special cases, fermionic and other sign problems can 
be eliminated with the loop-cluster algorithms 
\cite{chandrasekharan,cox,henelius}.

The new QMC methods have become important tools for quantum many-body 
research in condensed matter physics (with applications to quantum spins
\cite{harada,troyer,ssexy,ssespin1,ssespin2,wessel1,kato,sachdev,perc,wessel2,yunoki},
bosons \cite{hebert,dorneich1,schmid}, and one-dimensional fermion 
systems \cite{torsten,pinaki}) as well as in lattice gauge theory
\cite{chandrasekharan,cox}. An important property of some of the
loop-cluster algorithms is that they are efficient also in the presence of 
external fields \cite{kashurnikov,sse3,cox,syljuasen}. In particular the SSE 
algorithm with the operator-loop update \cite{sse3} has proven very 
powerful in several recent studies of quantum spin systems 
\cite{wessel2,yunoki} 
and boson systems \cite{hebert,dorneich1,schmid} including, respectively, a 
magnetic field and a chemical potential. It is interesting to note that in 
this respect QMC algorithms now perform better than classical Monte Carlo, 
since in the latter case external fields still pose challenging problems.

In this paper we present a general framework for constructing loop-type 
algorithms both in SSE and path integral methods. We focus primarily on the
SSE approach, which owing to the manifestly discrete nature of its
configuration space is easier to implement and, for the same reason, also 
is more efficient in most cases. In the SSE operator-loop update introduced 
in Ref.~\onlinecite{sse3}, a distinction was made between a general algorithm
(where it is necessary to allow the propagating end of the operator path to 
back-track) and special ones applicable only for certain Hamiltonians 
(where the paths do not back-track). For example, in the case of
the $S=1/2$ Heisenberg model with uniaxial anisotropy $\Delta$ and external 
magnetic field $h$ (also known as the XXZ-model), 
\begin{equation}
H = J\sum\limits_{\langle i,j\rangle}[ S_i^xS_j^x + S_i^yS_j^y
+ \Delta S_i^zS_j^z] -  h\sum\limits_{i} S^z_i,
\label{ham1}
\end{equation}
particularly efficient algorithms were devised at the isotropic Heisenberg
point ($\Delta=1$, $h=0$) and for the XY-model ($\Delta=0$, $h=0$). While the 
general algorithm can be used for any $\Delta,h$, it does not perform as
well in the limits $\Delta\to 1,h\to 0$ and $\Delta\to 0,h\to 0$ as the 
special algorithms exactly at these points (which are the only points at
which the more efficient algorithms can be used). Hence, one has to switch
algorithms when crossing the isotropic Heisenberg and XY points. The presence 
of such ``algorithmic discontinuities'' is clearly bothersome, both from a 
mathematical and practical point of view. Here we show how the algorithmic 
discontinuities can be eliminated within a more general framework of 
satisfying detailed balance when constructing the operator-loop. For reasons 
that will become clear below, we call the entities involved in this type of 
update ``directed loops''. With these, we are able to carry out simulations 
as efficiently in the limits approaching the Heisenberg and XY points as 
exactly at those points. We also show that this scheme can be easily 
adapted to continuous-time path integrals.

The outline of the rest of the paper is the following: In Sec.~II we review 
the SSE method and the operator-loop update on which the new directed loop
algorithm is based. We outline a proof of detailed balance and also discuss
a few special cases in which back-tracking can be easily avoided in the
loop construction. In Sec.~III we first discuss a more general condition for 
satisfying detailed balance in the SSE method, which leads us to the directed 
loop equations. We then show in detail how this scheme is applied to the 
spin-1/2 XXZ model. We present two solutions of the directed loop equations. 
One is identical to the previous generic operator-loop update and the other
smoothly connects to the special ``deterministic'' operator-loop 
algorithm at the isotropic Heisenberg point. We also briefly discuss the 
structure of the directed loop equations for a more general class of 
Hamiltonians. Implementation of directed loops in the path integral formalism 
is discussed in Sec.~IV. In Sec.~V we present simulation results in various 
parameter regions of the XXZ-model. We compare autocorrelation times for 
simulations using the two different directed loop solutions. We also extract 
the dynamic exponent in simulations of isotropic Heisenberg models at critical
points in one, two, and three dimensions. In Sec.~VI, as a demonstration of 
what can be accomplished with the improved solution, we present results for 
the magnetization as a function of the external field in the 2D Heisenberg 
model at very low temperatures. We calculate the magnetic susceptibility 
using gaps between different spin sectors extracted from the step-structure 
in the magnetization curve. We conclude with a summary and discussion in 
Sec.~VII. In an Appendix we outline the basic elements of a simple and 
efficient computer implementation of the SSE method.

\section{Stochastic series expansion}

The SSE method is a generalization \cite{sse1,sse2,sse3} of Handscomb's 
power series expansion method for the isotropic $S=1/2$ Heisenberg ferromagnet
\cite{handscomb} and antiferromagnet \cite{lyklema,lee} to a much wider range 
of systems. The performance is significantly improved also for the Heisenberg 
model \cite{sse4,ssespin1,ssespin2}. Early attempts of such generalizations 
\cite{chakravarty} were limited by the difficulties in analytically 
evaluating the traces of the terms of the expansion. This problem was solved 
\cite{sse1,sse2} by the development of a scheme for importance sampling also
of the individual terms of the traces expressed in a conveniently chosen 
basis. The starting point of the SSE method is hence the power series 
expansion of the partition function:
\begin{eqnarray}
Z =  {\rm Tr}\bigl\lbrace {\rm e}^{-\beta H} \bigr\rbrace
  =  \sum\limits_{\alpha}\sum\limits_{n=0}^\infty {(-\beta)^n \over n!} 
     \left \langle \alpha \left | H^n \right |\alpha \right \rangle ,
\label{zn1}
\end{eqnarray}
where the trace has been written as a sum over diagonal matrix elements in a
basis $\{|\alpha \rangle\}$. Simulation algorithms based on this expansion can
be formulated without sign problems for the same models as those for which 
world-line methods \cite{worldline} are applicable. There are no 
approximations causing systematic errors and very efficient loop-type updating 
algorithms have also recently been devised \cite{sse3,pinaki,athens,henelius2}.
A distinct advantage of SSE over continuous-time world-line methods 
\cite{prokofev,beard} is the discrete nature of the configuration space, 
which can be sampled without floating point operations.

Here we first review an implementation of the SSE method for the anisotropic 
$S=1/2$ Heisenberg model. A proof of detailed balance in the operator-loop 
updating scheme is then outlined. Several practical issues related to the 
operator-loops are also discussed. Estimators for physical observables will 
not be discussed here. Several classes of expectation values have been derived
in Ref.~\onlinecite{sse2}. Observables of interest in the context of the 
Heisenberg model have been discussed in Ref.~\onlinecite{sse4}. Off-diagonal 
correlation functions (single-particle Green's functions) have been 
studied in Ref.~\onlinecite{dorneich2}.

\subsection{SSE configuration space}

For the anisotropic Heisenberg antiferromagnet (\ref{ham1}) with $N$ spins
it is convenient to use the standard basis
\begin{equation}
|\alpha\rangle = |S^z_1,S^z_2,\ldots,S^z_N\rangle,
\end{equation}
and to write the Hamiltonian in terms of bond operators $H_b$, where
$b$ refers to a pair of sites $i(b),j(b)$,
\begin{equation}
H = -J\sum_{b=1}^{N_b} H_b, \hskip6mm (J>0).
\label{hbsum}
\end{equation}
For a d-dimensional cubic lattice the number of bonds $N_b=dN$. The bond 
operators are further decomposed into two operators;
\begin{equation}
H_b = H_{1,b} - H_{2,b},
\label{hbdef}
\end{equation}
where $H_{1,b}$ is diagonal and $H_{2,b}$ off-diagonal;
\begin{eqnarray}
H_{1,b} & = &
C - \Delta S^z_{i(b)}S^z_{j(b)} + h_b[S^z_{i(b)}+S^z_{j(b)}], 
\label{hb1} \\
H_{2,b} & = & \hbox{$1\over 2$}[S^+_{i(b)}S^-_{j(b)} + S^-_{i(b)}S^+_{j(b)}],
\label{hb2}
\end{eqnarray}
and we have defined the magnetic field on a bond; $h_b \equiv h/(2dJ)$. The 
constant $C$ should be chosen such that all matrix elements of $H_{1,b}$ 
are positive, i.e., $C \ge \Delta/4+h_b$. We will henceforth use the notation
\begin{equation}
C = C_0 + \epsilon,\hskip 5mm C_0 = \Delta/4+h_b,
\label{cdef}
\end{equation}
where $\epsilon \ge 0$. In the Hamiltonian (\ref{hbsum}) we have neglected 
a constant $N_bC$, which should be kept in mind when calculating the energy.

The powers of $H$ in Eq.~(\ref{zn1}) can be expressed as sums of 
products of the bond operators (\ref{hb1}) and (\ref{hb2}). Such a 
product is conveniently referred to by an operator-index sequence 
\begin{equation}
S_n = [a_1,b_1],[a_2,b_2],\ldots,[a_n,b_n],
\end{equation}
where $a_i \in \lbrace 1,2\rbrace$ corresponds to the type of operator 
($1$=diagonal, $2$=off-diagonal) and $b_i \in \lbrace 1,\ldots,N_b\rbrace$
is the bond index. Hence, 
\begin{eqnarray}
Z = \sum\limits_\alpha\sum\limits_{n=0}^\infty \sum_{S_n} 
     (-1)^{n_2}{\beta^n \over n!} 
     \left \langle \alpha \left | \prod_{i=1}^n H_{a_i,b_i} 
     \right | \alpha \right \rangle ,
\label{zn2}
\end{eqnarray}
where $\beta\equiv J/T$ and $n_2$ is the total number of spin-flipping 
operators $[2,b]$ in $S_n$. 
It is useful to define normalized states 
resulting when $|\alpha \rangle$ is propagated by a fraction of the SSE 
operator string: \begin{equation}
|\alpha (p)\rangle \sim \prod\limits_{i=1}^p H_{a_i,b_i} |\alpha\rangle .
\label{prop}
\end{equation}
Note that there is no branching, i.e., all $|\alpha (p)\rangle$ are basis 
states, and $|\alpha (p)\rangle$ and $|\alpha (p+1)\rangle$ are either the 
same state or differ only by a flipped pair of spins. In order for a term
$(\alpha,S_n)$ to contribute to the partition function the boundary condition 
$|\alpha (n)\rangle = |\alpha (0)\rangle$ has to be satisfied. On a bipartite 
lattice $n_2$ must therefore be even, and the expansion is then
positive definite. The terms (configurations) can thus be sampled using 
Monte Carlo techniques without sign problems. 

To simplify the Monte Carlo sampling, it is useful \cite{sse1} (although not 
necessary \cite{sse2}) to truncate the expansion at a maximum power $n = M$ 
and to insert $M-n$ ``fill-in'' unit operators $H_{0,0} \equiv 1$ in the 
operator products in all possible ways. This gives
\begin{equation}
Z = \sum\limits_\alpha \sum_{S_M} {\beta^n(M-n)! \over M!} 
    \left \langle \alpha  \left | \prod_{i=1}^M H_{a_i,b_i} 
    \right | \alpha \right \rangle ,
\label{zm}
\end{equation}
where $n$ now is the number of operators $[a_i,b_i] \not= [0,0]$. One can show
that {\cite{sse1,handscomb}} the average expansion order
\begin{equation}
\langle n \rangle = \beta N_b|E_b|,
\label{avn}
\end{equation}
where $E_b$ is the internal energy per bond, $E_b =-\langle H_b\rangle$ 
[including the constant $C$ in (\ref{hb1})], and that the 
width of the distribution is approximately $\langle n\rangle ^{1/2}$. $M$ can 
therefore be chosen so that $n$ never reaches the cut-off during the 
simulation ($M \sim \beta N$). The truncation error is then completely 
negligible. In practice, $M$ is gradually adjusted during the 
equilibration of the simulation, so that $M=a\times n_{\rm max}$, where 
$n_{\rm max}$ is the highest $n$ reached. A practical range for the factor 
$a$ is $1.2-1.5$. The simulation can be started with some random state 
$|\alpha \rangle$ and an ``empty'' operator string  $[0,0]_1,\ldots, [0,0]_M$ 
(we some times use the notation $[a,b]_p$ instead of $[a_p,b_p]$). Ergodic 
sampling of the configurations $(\alpha, S_n)$ is accomplished using two 
different types of updates.

\subsection{Updating scheme \label{section_oploop}}

The first update (diagonal update) is of the type $[0,0]_p \leftrightarrow 
[1,b]_p$, involving a single diagonal operator which changes the expansion 
order $n$ by $\pm 1$ \cite{sse4}. The corresponding Metropolis acceptance 
probabilities are
\begin{eqnarray}
P([0,0]_p \to [1,b]_p) & = & 
{N_b\beta \langle \alpha (p)| H_{1,b} | \alpha (p) \rangle \over M-n }, 
\label{diap1} \\
P([1,b]_p \to [0,0]_p) & = &
{M-n+1 \over N_b\beta \langle \alpha (p)| H_{1,b} | \alpha (p) \rangle },
\label{diap2}
\end{eqnarray}
where $P>1$ should be interpreted as probability one. The presence of $N_b$ in
these probabilities reflects the fact that there are $N_b$ random choices for 
the bond $b$ in a substitution $[0,0]\to [1,b]$ but only one way to replace 
$[1,b]\to [0,0]$ when $b$ is given. These diagonal updates are attempted 
consecutively for all $p=1,\ldots,M$, and at the same time the state 
$|\alpha \rangle$ is propagated when spin flipping operators $[2,b]$ are 
encountered (these cannot be changed in a single-operator update), so that 
the states $| \alpha (p) \rangle$ are available when needed to calculate 
the probabilities (\ref{diap1}) and (\ref{diap2}). 

The purpose of the second type of update --- the operator-loop \cite{sse3} --- 
is to accomplish substitutions $[1,b]_p \leftrightarrow [2,b]_p$ for a varying
number of operators, thereby flipping spins also in several of the propagated 
states (\ref{prop}). The expansion order $n$ does not change. It is then 
convenient to disregard the $[0,0]$ unit operator elements in $S_M$ and 
instead work with the original sequences $S_n$ of Eq.~(\ref{zn2}), which
contain only elements $[1,b]$ and $[2,b]$. For the discussion of the 
operator-loops, the propagation index $p$ will refer to this reduced 
sequence. It is also convenient to introduce two-spin states
\begin{equation}
|\alpha_{b_p}(p)\rangle = |S^z_{i(b_p)}(p),S^z_{j(b_p)}(p)\rangle ,
\label{alphap}
\end{equation}
i.e., the spins at bond $b_p$ in the propagated state $|\alpha (p)\rangle$ 
as defined in (\ref{prop}). The weight factor corresponding to (\ref{zn2}) 
can then be written as
\begin{equation}
W(\alpha,S_n) = {\beta^n\over n!}\prod\limits_{p=1}^n 
\langle \alpha_{b_p} (p) | H_{b_p} | \alpha_{b_p}(p-1) \rangle ,
\label{mproduct}
\end{equation}
where the non-zero two-spin matrix elements are
\begin{eqnarray}
\langle \downarrow \downarrow | H_b | \downarrow \downarrow \rangle &=&
\epsilon, \nonumber \\
\langle \downarrow \uparrow | H_b | \downarrow \uparrow \rangle &=&
\langle \uparrow \downarrow | H_b | \uparrow \downarrow \rangle =
\Delta /2 + h_b + \epsilon,  \nonumber  \\
\langle \uparrow \downarrow | H_b | \downarrow \uparrow \rangle &  = &
\langle \downarrow \uparrow | H_b | \uparrow \downarrow \rangle = 1/2,
\label{matrelem} \\
\langle \uparrow \uparrow | H_b | \uparrow \uparrow \rangle &=&
\epsilon + 2h_b .  \nonumber  
\nonumber
\end{eqnarray}
In principle the value of $\epsilon \ge 0$ is arbitrary but in practice a 
large constant is inconvenient since the average expansion order (\ref{avn})
has a contribution $\epsilon\beta N_b$. In many cases the simulation 
performs better with a small $\epsilon >0$ than with $\epsilon = 0$, however, 
as will be demonstrated in Sec.~V. For $\epsilon=0$, the number of allowed 
matrix elements is reduced from 6 to $4$ (if $h=0$) or 5 (if $h > 0$). 

\begin{figure}
\includegraphics[clip,width=8cm]{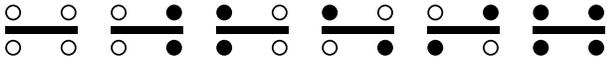}
\caption{The six different vertices corresponding to the matrix 
elements in Eqs.~(\protect{\ref{matrelem}}). The horizontal bar represents 
the full bond operator $H_b$ and the circles beneath(above) represent 
the spin state (solid and open circles for spin-$\uparrow$ and
spin-$\downarrow$, respectively) before(after) operation with either 
the diagonal or off-diagonal part of $H_b$.}
\label{vertices}
\end{figure}

The matrix element product in the weight (\ref{mproduct}) can be represented 
as a network of $n$ vertices, with two spins $S^z_{i}(p-1),S^z_{j}(p-1)$ 
``entering'' the $p$:th vertex and $S^z_{i}(p),S^z_{j}(p)$ ``exiting''. 
The six allowed vertices, corresponding to the non-zero matrix elements 
(\ref{matrelem}), are illustrated in Fig.~\ref{vertices}. The direction of 
propagation (here and in other illustrations) is such that moving upward 
corresponds to increasing the propagation  index $p$. 

\begin{figure}
\includegraphics[clip,width=6.0cm]{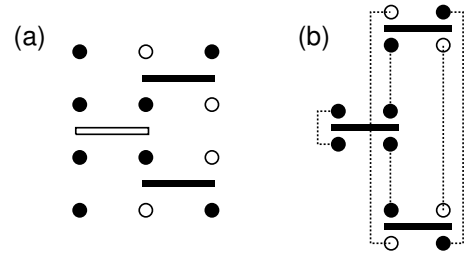}
\caption{(a) An SSE configuration for a three-site system with three 
operators, shown along with all the propagated states. Here open and solid 
bars indicate diagonal and off-diagonal operators, respectively. (b) The 
linked vertex list corresponding to (a). The dashed lines represent 
bidirectional links.}
\label{linked}
\end{figure}

In order to carry out the operator-loop update, a linked list of the vertices 
is first constructed. For each of the four legs on each vertex there is a 
spin state and a link to the following (in the direction of increasing $p$) 
or previous (direction of decreasing $p$) vertex leg at the same site. The 
periodic boundary condition of the propagated states must be taken into 
account, i.e., the links can span across $p=0$ and every leg then has an 
outgoing and incoming link (i.e., a bidirectional link). In case a spin (site)
is acted upon only by a single operator in $S_n$, the corresponding two legs 
of that vertex are linked together. Otherwise, for a site acted upon by two 
or more operators, all links are between different vertices. An example of 
an SSE configuration and its corresponding linked vertex list is shown in 
Fig.~\ref{linked}. Clearly, in an allowed configuration links can exist
only between legs in the same spin state. Note that in the representation 
with the full states in (a), which is never used in the actual simulation but
is included here for illustrative purposes, we distinguish between diagonal 
and off-diagonal operators (as is also done in the stored operator sequence
$S_M$ used in the diagonal update). In the vertex representation (b) the 
two-spin states are taken from the full propagated states (\ref{alphap})
and the type of the operator (diagonal or off-diagonal) is implicitly given 
by the four spin states. The bar is hence strictly redundant, but we include 
it in the figures as a reminder that the vertices represent matrix elements 
of the bond-operators.

\begin{figure}
\includegraphics[clip,width=6cm]{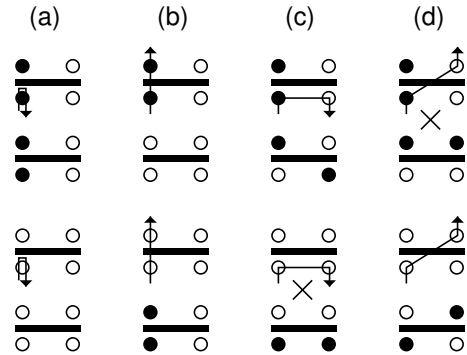}
\caption{All four paths through two vertices where the entrance is at the 
low-left leg. The arrow indicates the exit leg. The resulting updated vertices,
with the spin at the entrance and exit legs flipped, are also shown. The
two cases marked with an X are forbidden, since the updated vertices do 
not correspond to operators in the Hamiltonian considered here. We refer
to the four different processes as (a) ``bounce'', (b) ``continue-straight'',
(c) ``switch-and-reverse'', and (d) ``switch-and-continue''.}
\label{vpaths}
\end{figure}

\begin{figure}
\includegraphics[clip,width=7cm]{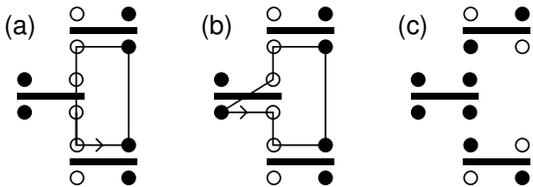}
\caption{Two different ways in which an operator-loop can close. The starting 
points of the loops in (a) and (b) are the legs from which the arrows point 
out. In (a) the last segment of the loop connects the initial and final 
vertices, resulting in the starting spin being flipped in the final
configuration. In (b) the last loop segment is within the initial vertex and 
the starting spin is flipped twice, with the net effect of no change. Both 
loops (a) and (b) here result in the updated vertices shown in (c).}
\label{closings}
\end{figure}

To construct an operator-loop, one of the $4n$ vertex legs is first selected 
at random as an initial entrance leg. One of the four legs belonging to the
same vertex as the entrance leg is then chosen as the exit from the vertex, 
and both the entrance and exit spins are flipped. Examples of how vertices 
change in the four types of process are shown in Fig.~\ref{vpaths}. The 
probability of exiting at a given leg, given the entrance leg and the four 
spin states defining the vertex, is taken proportional to that matrix element 
in (\ref{matrelem}) which corresponds to the vertex generated when the 
entrance and exit spins have been flipped. As an example, defining matrix 
elements obtained when flipping spins in a vertex as 
\begin{eqnarray}
& & W \left (_{f_1,f_2}^{f_3,f_4} \right ) (p) = \label{flipvtx} \\
& & \langle f_3S^z_i(p),f_4S^z_j(p)|H_{b}|f_1S^z_i(p-1),f_2S^z_j(p-1)\rangle,
\nonumber
\end{eqnarray}
where $f_i=-1$ if the spin on leg $i$ ($i=1,2,3,4$) is flipped and
$f_i=+1$ if it is not flipped, the probability of exiting at leg $2$ if
the entrance is at leg $1$ is given by
\begin{equation}
P_{2,1} = {W\left(_{--}^{++}\right ) \over
W\left(_{++}^{++} \right ) + 
W\left(_{--}^{++} \right ) +
W\left(_{-+}^{-+} \right ) +
W\left(_{-+}^{+-} \right) },
\label{exitprob}
\end{equation}
where we have used $\pm$ for $\pm 1$.
The reasons for this choice for the probability will be discussed in 
Sec.~II C. If the entrance and exit correspond to different 
sites (the switch-and-reverse and switch-and-continue processes in
Fig.~\ref{vpaths}), the change in the vertex corresponds to a change of the 
type of the operator (diagonal $\leftrightarrow$ off-diagonal). The leg to 
which the exit is linked is taken as the  entrance to the next vertex, from 
which an exit is again chosen. This procedure is repeated until the original 
starting point is reached (the loop closes). The mismatches (links connecting 
different spin states) existing at the original entrance and at the 
propagating end of the path are then ``healed'' and a new configuration 
contributing to the partition function has been created. Note that, depending 
on the way the loop closes, the spin at the leg from which the loop 
construction was started may or may not be flipped after the full loop has 
been completed. Examples illustrating this are given in Fig.~\ref{closings}. 

One of the two site-switching paths --- the switch-and-reverse in
Fig.~\ref{vpaths}(c) or switch-and-continue in ~\ref{vpaths}(d) --- 
is always forbidden since the corresponding off-diagonal matrix element 
of the Heisenberg bond-operator is zero. The bounce path in 
Fig.~\ref{vpaths}(a) is always allowed since the vertex is unaffected (the 
same spin is flipped twice, resulting in no net change). The 
continue-straight path \ref{vpaths}(b) is always allowed if the constant 
$\epsilon > 0$ so that all the diagonal matrix elements in (\ref{matrelem}) 
are larger than zero. If $\epsilon = 0$ at least one of the diagonal matrix 
elements vanishes, and the continue-straight process is then forbidden 
in some cases.

If a spin in the state $|\alpha \rangle$ is not acted upon by any of the
operators in $S_M$, it cannot be flipped by the operator-loop update.
Such ``free'' spins can, however, be flipped with probability $1/2$
since they do not appear in the weight function. Since the average of $n$, 
the number of operators in $S_M$, grows linearly with $\beta$, free spins 
appear frequently only at relatively high temperatures.

It is convenient to define a Monte Carlo step (MCS) as a sweep of diagonal
updates at all positions in $S_M$ where possible, followed by construction
of the linked list in which a number $N_l$ of operator-loops are constructed
before mapping back to a new $S_M$ and $|\alpha \rangle$ and flipping free
spins. Observables can be measured after every, or every few, MCS (in some 
cases, it may even worth-while to record measurements after every loop).

The remaining question now is how many operator-loops one should construct 
in each MCS. The operator-loops are typically of highly varying lengths. 
Each MCS should involve several loop updates, so that a significant fraction 
of the vertices are visited.  In order not to bias the measurements it is 
important that $N_l$ is fixed. One cannot, e.g., keep on constructing loops 
during a given MCS until the number of vertices visited exceeds a 
pre-determined number. The average size of the operator-loops depends strongly
on the model parameters. It is therefore useful to record the loop sizes and 
periodically adjust $N_l$ during the equilibration of the simulation. 
Typically, we determine $N_l$ such that the average cumulative loop length 
(the number of vertices visited) during one MCS is approximately 
$2\langle n\rangle$  or $2M$. In recording the loop length, we do not count 
bounces (since no change result in the vertex off which the path bounces). 
Among the counted steps there are still some fraction of back-tracking ones, 
i.e., segments of the operator-loop where completed vertex updates are 
reversed. If a bounce occurs already at the first step the loop closes 
immediately. With our definition, this is a completed loop of length $0$. In 
order not to bias the measurements, such length-0 loops also have to be 
counted among the $N_l$ completed loops. 

One could also fix $N_l$ based on a criterion involving the average number of 
leg-spins which are actually flipped during an MCS, but recording this 
number is slightly more complicated than just keeping track of the loop 
lengths. Since this has to be done only during equilibration the cost
is not prohibitive, however. The exact definition of $N_l$ and precisely what 
constitutes one MCS are not critical issues (as in the classical Wolff cluster
algorithm \cite{wolff}, where the MCS can also be defined in a way analogous 
to what we have discussed here).

The operator-loop construction (the operator path) is a type of random 
walk in a d+1-dimensional space (although the network of connected vertices 
does not necessarily have this dimensionality---it could effectively have a 
fractal dimension $< d+1$). One may therefore wonder whether the closing of 
the loop could become problematic, especially for large systems in three 
dimensions. In some cases, an operator-loop can indeed become very long before
it closes. In rare cases a loop may even not close during a simulation of 
practical length. The loop size distribution is always very broad, however,
and the non-closing problem can simply be circumvented by imposing a maximum 
length beyond which the loop construction is terminated. The way we typically 
implement this termination is by immediately initiating a new MCS (beginning
with a diagonal update), hence disregarding all the loops that were 
constructed during the MCS of the terminated loop. This way, we do not have 
to save actual operator paths (needed in order to undo the 
changes done during construction of the terminated loop), which would become 
impractical for long paths. The termination does not 
violate detailed balance and hence the correct distribution of configurations 
contributing to $Z$ is maintained. Termination of incomplete loops does 
introduce a bias in quantities which are related to the extended configuration
space, however, such as single-particle Green's functions \cite{dorneich2}. 
Typically, we use a maximum loop length $\approx 100\times \langle n\rangle$. 
For the XXZ-model (in any number of dimension) incomplete loop termination 
is then extremely rare (excessively long loops can occur more frequently 
in other models \cite{pinaki}). The average loop length is typically much 
smaller than $\langle n\rangle$, but can in some cases be a significant 
fraction of $\langle n\rangle$ (up to tens of percent).

\subsection{Detailed balance}

In the originally proposed operator-loop scheme \cite{sse3}, the probability
of selecting an exit leg is proportional to the corresponding matrix element 
(\ref{matrelem}) when the entrance and exit spins have been flipped (with 
the factor of proportionality chosen to give probability one for the sum of 
the four probabilities), a specific example of which is written in 
Eq.~(\ref{exitprob}). One can prove that detailed balance is satisfied in 
this process by considering an extended configuration space which includes 
also the intermediate configurations generated during the loop construction 
(which do not contribute to the partition function). 

\begin{figure}
\includegraphics[clip,width=6.5cm]{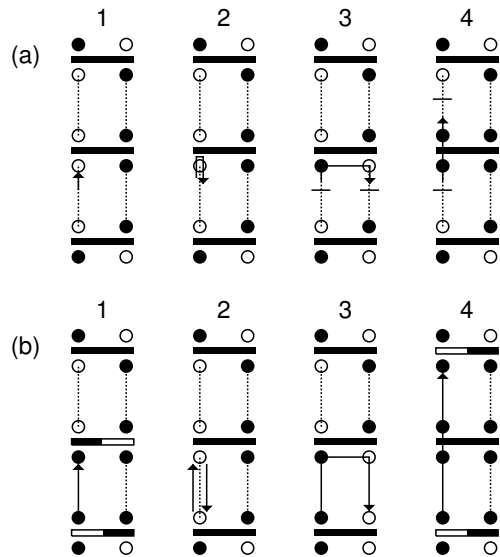}
\caption{Two ways to look at the extended configuration space generated
during operator-loop construction. Examples of how the configuration shown in
(a)-1 is modified at the beginning of an update in the link-discontinuity 
picture (a) and ladder operator picture (b) are shown. In (a), the arrow in 
1 indicates the proposed starting point of the loop. In (b), a first step of 
flipping the two spins at this link has already been carried out 
(generating ladder operators which are indicated by vertices 
with semi-filled bars), and the arrow indicates the entrance point for the 
following step. In both (a) and (b), configurations that can be generated 
out of 1 are shown in 2-4. Link-discontinuities are indicated by small
horizontal lines in (a). In both cases, configuration 2 corresponds to the 
bounce process, which results in immediate return to the original 
configuration.}
\label{lpictures}
\end{figure}

The detailed balance proof is 
illustrated by an example for a configuration with three vertices in 
Fig.~\ref{lpictures}(a). In 1), the leg with the arrow has been selected as 
the initial entrance point of the operator-loop. An exit leg is chosen 
according to the probabilities discussed above. 
Flipping both the entrance spin and the exit spin leads to a new configuration
in the extended space. In Fig.~\ref{lpictures}(a) the three resulting 
configurations which have non-zero probability are shown in 2)-4). The 
entrance $\rightarrow$ exit paths are also indicated and the corresponding 
spins have been flipped. The probability of process 3) corresponds to the 
example given in Eq.~(\ref{exitprob}), which when inserting the actual spin 
states becomes
\begin{equation}
P_{2,1} = {\langle \uparrow\downarrow|H_b|\downarrow\uparrow\rangle
\over \langle \downarrow\uparrow|H_b|\downarrow\uparrow\rangle +
\langle \uparrow\downarrow|H_b|\downarrow\uparrow\rangle +
\langle \uparrow\uparrow|H_b|\uparrow\uparrow\rangle }.
\end{equation}
In 2), the entrance and exit are at the same leg. This is a bounce process 
which closes the loop immediately with no change in the configuration. 
In 3) and 4) the vertex has changed and two links have 
appeared which connect legs with different spin states. We call these 
``link-discontinuities''. Only configurations with no link-discontinuities 
contribute to $Z$. All configurations created during the loop construction 
contain two link-discontinuities, until the loop closes (which can be seen as 
the discontinuities annihilating each other). There are no weight changes 
associated with the link-discontinuities---the configuration weight is still 
considered to be given by Eq.~(\ref{mproduct}). Hence, the only weight change 
arises from the change in the affected vertex when the entrance and exit 
spins are flipped. 

The way the exit leg is chosen at the start of the operator-loop corresponds 
to a heat-bath algorithm. The probabilities of no change (staying in the 
original subspace) or transfer to a configuration with two 
link-discontinuities are proportional to the respective weights in the 
extended space. Once a configuration with two discontinuities has been 
created (i.e., the first step was not a bounce), we do not want to create more 
discontinuities (which would take us out of the extended space considered 
here) and therefore the following updates can only take place at the 
discontinuities (the end points of the path), i.e., the discontinuities can 
be moved. Here the same heat-bath algorithm as in the first step is used. 
The only difference is that the entrance leg is not chosen at random but is 
given by a link from the previous vertex. Hence, the whole process consists 
of a series of heat-bath steps, which satisfy detailed balance and therefore 
generate configurations according to probabilities proportional to the weight 
in the extended space. The subset of configurations with zero 
link-discontinuities, which contribute to $Z$, are therefore also generated 
with the correct distribution. The process is ergodic because all types of 
vertices can be generated, and since the operator path can wind around the 
periodic boundaries and then change both the spatial winding number and the 
total magnetization. Within a sector of fixed winding number and 
magnetization, local updates which constitute a small subset of the 
operator-loops suffice to ensure ergodicity \cite{sse4}.

Instead of thinking about the extended configuration space in terms of
link-discontinuities, one can consider the vertices created when one of 
the spins in the original vertices of Fig.~\ref{vertices} is flipped. 
These new vertices correspond to the single-spin flipping (ladder) 
operators $S^+_i$ and $S^-_i$. The loop construction can be formulated
in terms of introducing pairs of these, which are then randomly propagated 
until they reach the same vertex and annihilate each other. The start of such
a process is illustrated in Fig.~\ref{lpictures}(b), using the same 
configuration and starting point as in \ref{lpictures}(a). The difference 
with respect to the previous discussion is that now there are no 
link-discontinuities. Instead, the spins at both ends of the link 
at the selected entrance leg are flipped simultaneously. 
This introduces two ladder vertices. Here one has to assign a value, $v_l$, 
to the matrix elements of the ladder operators (i.e., the new operators 
are $v_lS^+_i$ and $v_lS^-_i$). The initial loop segment, an example of which 
is shown as 1) in Fig.~\ref{lpictures}(b), is then generated only with a 
probability min$[1,v_l^2/(W_1W_2)]$, where $W_1$ and $W_2$ are the matrix 
elements corresponding to the two vertices that are considered for replacement 
by ladder vertices. If this first step is accepted the next step is again to 
choose an exit leg. As before, the propagation of the path is carried out 
according to a heat-bath algorithm, with probabilities proportional to the 
matrix element when the entrance and exit spins have been flipped. In the 
example, paths 2) and 3) lead to closed loops (back to the space with no ladder
operators), whereas in 4) the ladder operators are moved further away from 
each other. Note that both spins on the link corresponding to the exit leg 
are flipped in every steps, so that no link-discontinuities appear. The 
process continues until the two ladder operators are on the same vertex, 
which then becomes equal to one of the original two-spin vertices. This 
brings the system back into the original configuration space.

The link-discontinuity and ladder operator pictures of the loop construction 
are clearly completely equivalent, although the probabilities associated with 
starting (or closing) the loop are different. In actual simulations it is 
typically more convenient to use the link-discontinuities view. The ladder 
operator picture explicitly relates the extended configuration space to that 
of correlation functions involving these operators, but the 
link-discontinuities can be easily related to them as well. An extended 
configuration space analogous 
to the one generated in the operator-loop update was first utilized in the 
context of the worm algorithm for continuous-time world-line simulations 
\cite{prokofev}. The issue of measuring off-diagonal correlation functions 
using the SSE operator-loops has been considered in Ref.~\cite{dorneich2}. 

In Sec.~III we will give a more formal and complete proof of detailed balance. 
We will show that the heat-bath algorithm is not the only, and also not the 
most efficient, way to satisfy detailed balance when constructing the 
operator-loop. We will introduce the concept of a directed loop to form 
a general framework for loop updating schemes, both in SSE and path integral 
simulations. In the SSE scheme, the directed loop simply leads to different
probabilities of choosing among the four exits from a vertex, all other
aspects of the method remaining as has been discussed in this section.
Before introducing the directed loop concept, we first consider 
special cases in which the bounce process can be excluded. In II E we
discuss more technical SSE implementation issues, which also are common
to the heath-bath operator loops and the new directed loops.
 
\subsection{Excluding back-tracking in special cases}

In the general operator-loop algorithm discussed above, the probability of 
the bounce process is always non-zero, because  the vertex remains unchanged 
and has a non-zero value (otherwise, it would not appear in the configuration
in the first place). In some special cases, it is possible to modify the 
algorithm in such a way that the bounce is completely excluded. This has
very favorable effects on the simulation dynamics, since there is then no 
back-tracking and all segments of the loop accomplish changes in the 
configuration. 

\begin{figure}
\includegraphics[clip,width=6.5cm]{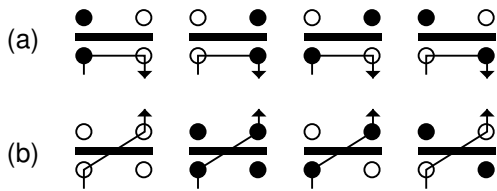}
\caption{All allowed vertices in the deterministic operator-loop algorithm
in the case of the Heisenberg antiferromagnet (a) and ferromagnet (b). 
Operator-loop segments starting at the lower left leg are also shown.}
\label{dpaths}
\end{figure}

The most important of the special cases is the isotropic Heisenberg model 
($\Delta=1,h=0$) \cite{sse3}. A very similar algorithm exists for the
ferromagnet ($J < 0$) \cite{henelius}. For the antiferromagnet, choosing the 
constant $\epsilon=0$ in Eqs.~(\ref{matrelem}) implies that the vertices with 
all spins up or all spins down vanish and the remaining four matrix elements 
all equal $1/2$. As a result, the matrix element product in (\ref{mproduct}) 
is simply $(1/2)^n$ and is not affected by the operator-loop update. If the 
bounce process is excluded, the only remaining process is the 
switch-and-reverse shown in Fig.~\ref{dpaths}(a) and the path is hence 
deterministic. The actual loop structure is only changed by the 
diagonal update. The deterministic 
loop process is clearly symmetric with respect to flipping or flipping back 
the spins at all vertex legs covered by the loop  and hence it obeys 
detailed balance. For the ferromagnet, the bounce can be excluded 
if $C=-1/4$ in (\ref{hb1}) [for the isotropic ferromagnet $\Delta
= -1$ and there is no minus sign in (\ref{hbdef})], and the only remaining 
process is then the switch-and-continue process shown in 
Fig.~\ref{dpaths}(b).
 
In the deterministic case, each vertex leg can be uniquely assigned to a loop,
and the loops can be flipped 
independently of each other. Instead of randomly choosing starting points 
and constructing a fixed number of loops, one can then construct all possible 
loops exactly once, by always picking a starting point which does not belong 
to a loop already constructed. The loops should then be flipped with 
probability $1/2$. The random decision of whether or not to flip can be made 
before the loop is constructed, but even if the decision is not to flip one 
has to construct the whole loop and set flags on the vertex legs visited, so 
that one does not attempt to construct the same loop again. Loops are 
constructed this way until all $4n$ vertex legs have been visited. This method
of constructing all the loops is analogous to the classical Swendsen-Wang 
multi-cluster method \cite{swendsen}, whereas, as was already mentioned above,
the operator-loop construction in the general non-deterministic case is more 
similar to the Wolff single-cluster algorithm \cite{wolff}. 

It should be 
noted that in the deterministic case an algorithm including only operator
updates (diagonal updates and loops) is not completely ergodic. In the
antiferromagnet, spin states with all spins up or down are isolated from the 
other states since no operators can act on them. These two states are 
important only at very high temperatures and they can then be reached by also 
performing random flips of ``free'' spins. In simulations with $\epsilon > 0$ 
all states can be reached even without such spin flips.

Another special case is the XY-model \cite{sse3,ssexy} ($\Delta=0$, $h=0$). 
In this case all matrix elements in (\ref{matrelem}) equal $1/2$ if
the constant $\epsilon=1/2$. The weight is then again only dependent on $n$ 
and does not change in the operator-loop update. The bounce can therefore be 
excluded also in this case, leaving two remaining allowed exits from each 
vertex. Although these paths are not deterministic, one can still subdivide 
the system into loops that can be flipped independently of each other.

The loop structure in the general operator-loop algorithm, which includes
bounce processes, is similar to that in the worm algorithm for continuous-time 
path-integrals \cite{prokofev}, although the two methods are quite different 
in other respects (the actual processes used to construct the SSE 
operator-loops and the worms are different---see Sec.~VII). In the special 
cases where the bounce process can be excluded, the SSE operator-loops are 
analogous to the world-line loops (in discrete \cite{evertz} or continuous 
\cite{beard,syljuasen} imaginary time). The close relationships between the 
Euclidean path integral in continuous time and the discrete representation 
on which the SSE method is based has been discussed in previous papers
\cite{sse2,irsse,athens} and will also be further elucidated here in Sec.~IV. 

\section{Directed Loops \label{DLsection}}

In the operator-loop update discussed in the previous section, detailed 
balance is satisfied using a heat-bath algorithm for propagating the path 
between connected vertices. In this section we will present a more general set
of equations that have to be satisfied for detailed balance to hold in such a 
process. We will show that these equations have an infinite number of 
solutions, some of which can lead to a more efficient sampling than the heat 
bath. We construct a particular solution based on the intuitive hypothesis 
(for which we have no rigorous proof) 
that the probability of bounces (back-tracking) 
should be minimized. We show that the bounces can in fact be completely 
excluded in a much wider range of parameters than at the two isolated points 
(isotropic XY and Heisenberg) discussed in Sec.~II D. 

We call the entities involved in the more general scheme ``directed loops'', 
because the detailed balance equations that we construct (the directed loop
equations) explicitly take into account the fact that the construction of 
the path of vertices is directional, i.e., the probability of exiting at a 
particular leg given the entrance leg is not the same as the probability 
of the reverse process. The original operator-loop update with the heat-bath 
probabilities \cite{sse3} discussed in the previous section corresponds to a 
particular solution of the directed loop equations. We stress that if another
solution is used, the only difference in the actual simulation with respect 
to the original scheme is a different set of probabilities for exiting at a 
given vertex leg, given the entrance leg and the four spin states. Before we 
explicitly construct new solutions in the context of the XXZ-model we 
begin by describing more generally how the directed loop equations arise.

\subsection{Conditions for detailed balance}

Let us first recall that the detailed balance requirement reads
\be
	P(s \to s^\prime) W(s) = 
	P(s^\prime \to s) W(s^\prime),
\label{detbalance}
\ee
where $s$ denotes a configuration having weight $W(s)$, which in the SSE 
method is expressed as a product over vertex weights, Eq.~(\ref{mproduct}), 
and $P(s \to s^\prime)$ is the probability of changing the 
configuration from $s$ to $s^\prime$. While the weights are given by the 
Hamiltonian the probability for how to update the spin configuration 
depends on the actual algorithm used. 

The algorithm for constructing an operator-loop to update an SSE 
configuration is quite general for any form of the 2-body interaction
(and can be extended also to multi-particle interactions). With the
configuration mapped onto a linked vertex list, an initial entrance 
vertex leg is first picked at random among all $4n$ legs. Then
an exit leg belonging to the same vertex is chosen in a probabilistic way 
and the spins on the entrance and exit legs are flipped with unit 
probability. More generally, the states at these legs are updated with
non-zero probabilities only for changes leading to vertices corresponding
to non-zero matrix elements. For simplicity, we here assume that the
change at the exit leg is uniquely dictated (through conservation laws) by 
the change at the entrance leg. The process continues using as the new 
entrance leg the leg linked to the exit leg. The process stops when the 
initial starting leg is reached. The probability for arriving at a new 
configuration $s^\prime$ can therefore be written as
\bea
	P(s \to s^\prime) & = & \sum P(e_0) P(s,e_0 \to s_1,e_1) 
	\nonumber \\ 
                    	  &   & \times P(s_1,e_1 \to s_2,e_2)
	                        \cdots \label{updateprob} \\
	                  &   & \times P(s_{n-1},e_{n-1} \to s^\prime,e_0),
	\nonumber
\eea
where $P(e_0)$ is the probability for choosing the vertex leg $e_0$ as
the initial starting point and $P(s_i,e_i \to s_{i+1},e_{i+1})$ is
the probability given a spin configuration $s_i$ and the entrance leg
$e_i$ to exit the vertex at $x_i$ which is connected to the next entrance
leg $e_{i+1}$, resulting in a new configuration $s_{i+1}$. The intermediate
configurations $s_i$ belong to the extended space of configurations with
two link-discontinuities, as discussed in Sec.~II-C. The exit legs $x_i$
do not explicitly appear in the probabilities since we have assumed that 
they are uniquely linked to the following entrance legs $e_{i+1}$ 
(generalization to cases where the uniqueness does not hold are 
straight-forward). The sum is over all possible closed 
loops which result in the updated configuration being the particular 
configuration $s^\prime$. To find a convenient way of choosing the 
probabilities on the right hand side of the above one needs an
expression for the inverse process where the spin configuration 
$s^\prime$ is transferred into $s$. This can be written down quite simply 
by realizing that for each of the terms in Eq.~(\ref{updateprob}) 
there is a corresponding term which describes the ``time''-reversed 
path, which contributes to the reverse probability. Thus one can write
\bea
	P(s^\prime \to s) & = & \sum P(e_0)P(s^\prime,e_0 \to s_{n-1},e_{n-1}) 
	\nonumber \\ 
                &   & \times \cdots \times P(s_2,e_2  \to s_1,e_1) \\
	         &  & \times P(s_1,e_1 \to s,e_0), \nonumber 
\eea
where the sum is over the {\em same} closed loops as in Eq.~(\ref{updateprob}).
By inserting these expressions into the detailed balance equation
(\ref{detbalance}) we see that detailed balance is satisfied if
\bea
	& & W(s_i) P(s_i,e_i \to s_{i+1},e_{i+1}) = \nonumber \\
	& & W(s_{i+1})P(s_{i+1},e_{i+1} \to s_i,e_{i}) \label{detbalcrit}
\eea
for all possible SSE configurations and entrance legs. Because the update
$(s_i,e_i \to s_{i+1},e_{i+1})$ changes only one particular vertex, all except
one of the factors in the product of vertex weights in Eq.~(\ref{mproduct}) 
factor out and cancel. Writing $W(s,e,x) = W_s P(s,e \to s^\prime,x)$, 
where we have slightly changed the notation so that $W_s$ denotes the matrix 
element corresponding to a {\em single} vertex with its four leg states 
coded as $s$, $e$ is the entrance leg, and $x$ is the exit leg on the same 
vertex, one can formulate the detailed 
balance criterion Eq.~(\ref{detbalcrit}) as
\be
	W(s,l_1,l_2) = W(s^\prime,l_2,l_1) \label{constraint}
\ee	  
which should be valid for all possible vertex types which can be converted 
into 
each other by changing the states at the entrance and exit legs. This equation
implies many relations between the unknown probabilities of how to chose an 
exit leg given a particular vertex and an entrance leg. There are additional 
relations which must be satisfied. Requiring that the path always continues 
thru a vertex translates into into 
\be
	\sum_{x} P(s,e \to s_x,x) = 1,
\ee
where the sum is over all legs on the vertex. We have emphasized in the
notation $s_x$ that the resulting spin configuration depends on the 
exit leg. In terms of the weights $W(s,l_1,l_2)$ this requirement translates 
into
\be \label{sumprob}
	\sum_{x} W(s,e,x) = W_s,
\ee
which must be valid for all vertices and entrance-legs. This set of equations, 
Eq.~(\ref{sumprob}) together with the relations in Eq.~(\ref{constraint}), 
form the directed loop equations, the foundations of our new approach to
construct valid probability tables for the operator-loop update.

\subsection{SSE directed loops for the XXZ-model}

\begin{figure}
\includegraphics[clip,width=3.5cm]{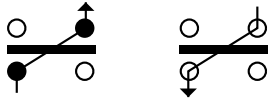}
\caption{Example of two vertices with directed loop segments 
that transform into each other in the 
flipping process.\label{related}}
\end{figure}

\begin{figure}
\includegraphics[clip,width=8cm]{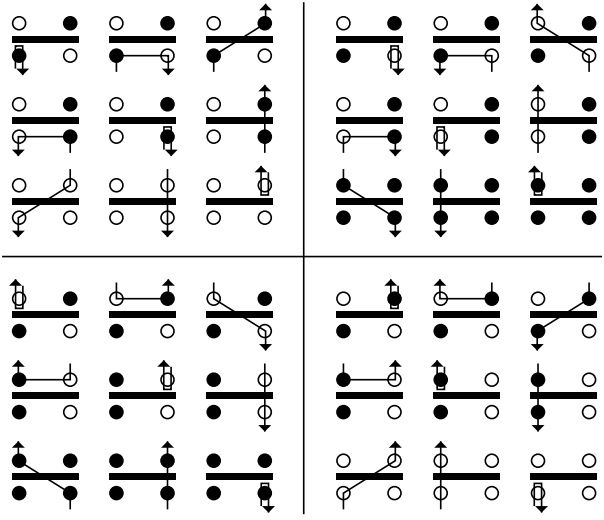}
\caption{Possible assignments of directed loop segments for half of the 
different combinations of vertices and entrance-legs. The other half of the 
vertex configurations can be obtained by interchanging up and down spins (solid
and open circles) while keeping the arrows. The lines with arrows are the 
directed loop segments. The configurations are divided into four sets 
(one in each quadrant). On flipping the spins connected by the loop segment 
and reversing the direction of the arrow, only configurations within the 
{\em same} quadrant are transformed into each other.
\label{all}}
\end{figure}

For the XXZ-model there are just three possible exits for any given entrance 
leg as one  choice always leads to a zero weight state when spins connected by
the loop segment are flipped (due to violation of the $z$-magnetization 
conservation of the model). Fig.~\ref{vpaths} illustrates the possibilities 
for placing directed loop segments for different vertices. 
In order for our update process to satisfy detailed balance we recall that
according to Eq.~(\ref{constraint}) we must relate vertices 
in which the two spins connected by the 
loop segment are flipped {\it and} the direction of the loop segment is 
reversed. Such related configurations are illustrated in Fig.~\ref{related}. 
Furthermore Eq.~(\ref{sumprob}) relates vertices with different
exit legs having the same spin configuration and entrance legs. We then make 
the key observation that all possible vertex configurations
can be divided into eight subsets that do not transform into each other. 
Half of these subsets are shown in Fig.~\ref{all}, where only configurations 
within the {\em same} quadrant are transformed into each other. 
These configurations form closed sets under the flipping operation. It is 
therefore sufficient to derive the detailed balance conditions, 
Eq.~(\ref{constraint}), for transitions between vertex
configurations in the same set independently of other configurations.

A row in any of the quadrants in Fig.~\ref{all} contains all three
configurations which can be reached by entering a certain vertex 
from a certain entrance leg. For instance, in the upper left quadrant the 
entrance leg for the first row is the lower left one, for the second row
the lower right one, and for the third row the upper right one. According to
Eq.~(\ref{sumprob}), summing the weights of all possible 
configurations that can be reached from a certain in-leg, keeping the 
spin configuration fixed, should equal the vertex weight alone. Thus taking 
the upper left quadrant of Fig.~\ref{all}, we have for rows 1-3 from the top:
\bea 
	W_1 & = & b_1 + a   + b, \label{detbal1} \nonumber \\
	W_2 & = & a   + b_2 + c, \label{detbal2} \label{im123} \\
	W_3 & = & b   + c + b_3, \label{detbal3} \nonumber
\eea
where the symbols on the left hand sides are the vertex weights 
Eq.~(\ref{matrelem}) in the spin configuration space, i.e.,
\bea
W_1 &=& \langle \uparrow \downarrow | H_b | \downarrow \uparrow \rangle = 
\langle \downarrow \uparrow | H_b | \uparrow \downarrow \rangle = 1/2,
\nonumber \\
W_2 &=& \langle \downarrow \uparrow | H_b | \downarrow \uparrow \rangle =
\langle \uparrow \downarrow | H_b | \uparrow \downarrow \rangle =
\Delta /2 + h_b + \epsilon,  \nonumber  \\
W_3 &=& \langle \downarrow \downarrow | H_b | \downarrow \downarrow \rangle =
\epsilon, \label{w1234} \\
W_4 &=& \langle \uparrow \uparrow | H_b | \uparrow \uparrow \rangle =
\epsilon + 2h_b ,  \nonumber  
\eea
while those on the right are weights in the enlarged 
configuration space of spins and directed loop segments. We have assigned 
equal weights to the configurations which are related by flipping, in 
accordance with Eq.~(\ref{constraint}). The order of the symbols on the 
right-hand sides of Eqs.~(\ref{im123}) follows the order in the upper left 
quadrant of Fig.~\ref{all}, so that, e.g., the weight of the two 
configurations in Fig.~\ref{related} is $b$ and the weight of the very 
middle configuration in the upper left quadrant of Fig.~\ref{all} is $b_2$. 
We use $b$ with a subscript to denote a weight of a configuration where 
the exit equals the entrance (bounce process).

As mentioned above there are in all eight sets of vertex configurations which
close under the flipping process. These sets are in principle independent 
of each other and have their own equation sets. However, one can easily 
convince one-self that because of symmetry reasons there are only two 
different types of sets. One of these symmetries is that of permuting the 
two spins acted upon by $H_b$. This implies that the equations derived for 
the set in the upper left quadrant in Fig.~\ref{all}, Eqs.~(\ref{im123}), 
are the same as for the set (not shown) that can be obtained from the upper 
right quadrant by interchanging up- and down spins, keeping the orientation 
of the directed loop segments.
The other symmetry is that of imaginary time inversion, which in the figures 
corresponds to switching the pairs of spins below and above the horizontal 
bar representing the operator $H_b$. This symmetry together with the previous
one identify the rules for the upper left quadrant of Fig.~\ref{all} with 
those of the lower right quadrant. Thus, one only has to consider two 
independent sets of equations, Eqs.~(\ref{im123}) and the corresponding 
equations which can be derived from the lower left quadrant in Fig.~\ref{all}:
\bea 
W_1 & = & b^\prime_1+a^\prime+b^\prime,  \label{detbal4} \nonumber \\
W_2 & = & a^\prime+ b^\prime _2+c^\prime, \label{detbal5} \label{im456} \\
W_4 & = & b^\prime+c^\prime+b^\prime _3. \label{detbal6}  \nonumber
\eea
This latter set takes the form of the set (\ref{im123}) but with primed 
symbols to denote the weights and $W_4$ instead of $W_3$. There is a further 
reduction in the case of zero magnetic field, where the two equation sets 
become identical.  

Before discussing solutions to these sets of equations it should be stressed
that the actual probabilities for selecting the exit leg is given by dividing 
the weight in the extended configuration space by the weight of the bare
vertex, so that, e.g., the probability for choosing the ``bounce'' process 
given that the entrance leg is the lower left one on a vertex with weight 
$W_1$, as shown in the very upper left hand corner of Fig.~\ref{all}, 
is $b_1/W_1$. 

It is clear that there are many solutions with only positive weights to the 
above equation sets as they are under-determined; both sets have six 
unknowns and there are three equations with the additional requirement of 
non-negative weights. A particular symmetric solution is the one corresponding
to the heat-bath probabilities used in the original scheme \cite{sse3}, 
which we will henceforth refer to as Solution A. It is given by
\bea
	a   & = & W_1 W_2/(W_1 + W_2 + W_3), \nonumber  \\
	b   & = & W_1 W_3/(W_1 + W_2 + W_3), \nonumber \\
	c   & = & W_2 W_3/(W_1 + W_2 + W_3), \label{symmetricsol} \\
	b_i & = & W_i^2/(W_1 + W_2 + W_3). \nonumber
\eea
For the primed weights, $W_3$ is replaced by $W_4$. Clearly the probabilities 
for choosing the exit leg are here proportional to the weights of the resulting
bare vertices which are obtained by flipping the two spins on the loop 
segment, an example of which was given in Eq.~(\ref{exitprob}).
This solution is valid in the full parameter space of the XXZ-model. 
However, it generally assigns a relatively large weight to the bounce 
processes where the exit leg equals the entrance leg. These are ineffective 
in updating the configurations. In particular, when the field $h\to 0$
and the anisotropy $\Delta \to 1$ the bounce probability approaches $1/2$. 
Although the method still is reasonably efficient (we are not aware of any 
method that has been more successful for models including external fields), 
this is bothersome since the SSE algorithm exactly at $h=0$ can be formulated 
entirely without any bounce processes \cite{sse3}, as reviewed in Sec.~II D,
and is then considerably more efficient. The fact that the $h=0$ scheme has 
no bounces and is completely deterministic, whereas the $h \to 0$ method has 
bounce probabilities approaching $1/2$, inspires us to look for solutions 
where the bounce probability instead vanishes continuously as $h \to 0$. 
This will eliminate the ``algorithmic discontinuity'' of the previous approach.

\begin{figure}
\includegraphics[clip,width=8.4cm]{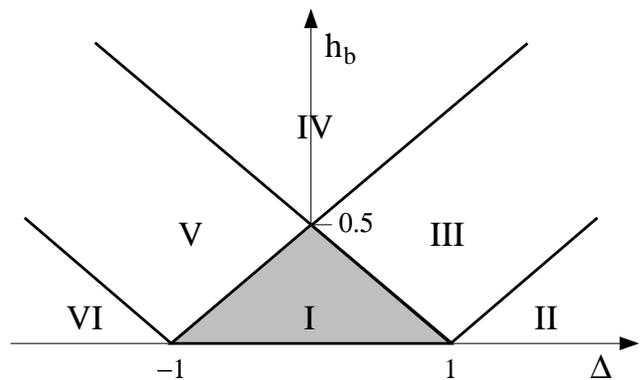}
\caption{``Algorithmic phasediagram'' showing regions where various bounce 
weights must be non-zero. The actual values of these weights are given in
Table I. In the shaded region all bounce weights can be set to zero. Other
bounce weights are listed in Table I. 
\label{phasediagram}}
\end{figure}

For the discussion of other solutions to the directed loop equations 
(\ref{im123}) and (\ref{im456}) it is convenient to express these equations
in terms of the bounce weights $b_1,\ldots,b^\prime_3$:
\bea
a & = & \f{1+\Delta}{4} + \f{h_b}{2} + \f{-b_1-b_2+b_3}{2}, \nonumber\\
b & = & \f{1-\Delta}{4} - \f{h_b}{2} + \f{-b_1+b_2-b_3}{2}, \nonumber \\
c & = & \f{\Delta-1}{4} + \f{h_b}{2} + \epsilon + \f{b_1-b_2-b_3}{2}, 
\label{dim123456} \\
a^\prime & = & \f{1+\Delta}{4} - \f{h_b}{2} + 
\f{-b^\prime_1-b^\prime_2+b^\prime_3}{2}, \nonumber \\
b^\prime & = & \f{1-\Delta}{4} + \f{h_b}{2} + 
\f{-b^\prime_1+b^\prime_2-b^\prime_3}{2}, \nonumber \\
	c^\prime & = & \f{\Delta-1}{4} + \f{3h_b}{2} + 
\epsilon+ \f{b^\prime_1-b^\prime_2-b^\prime_3}{2}, \nonumber
\eea
where we have explicitly inserted the expressions for the vertex weights, 
Eq.~(\ref{w1234}). We seek positive solutions to these equations. Being 
under-determined there are many solutions, so we will try to find the 
solutions that yield the most effective algorithms. As a general principle 
for finding efficient rules, we will attempt to minimize the bounce weights 
$b_1,\ldots,b^\prime_3$.
The solution so obtained will be termed Solution B. Inspecting the equations, 
it is clear that there is one region in parameter space where one can avoid 
bounces altogether. This region is shown as the shaded region in 
Fig.~\ref{phasediagram}. From the requirement of non-negative vertex weights 
we already have the restriction $\epsilon \geq 0$. In the shaded region, the 
requirement of non-negative weights also in the enlarged configuration 
space when all the bounce weights are zero imposes an additional constraint
on $\epsilon$: $\epsilon \geq (1-\Delta)/4 -h_b/2$. We have no rigorous
principle of finding the optimal value of $\epsilon$ in general, but as can 
be inferred from our simulation tests (presented in Sec.~V) it is often 
advantageous to choose a small but non-zero value in
cases where $\epsilon_{\rm min}=0$. 

For the Heisenberg antiferromagnet at zero magnetic field ($\Delta=1,h=0$)
the deterministic algorithm constructed in Ref.~\onlinecite{sse3} is
recovered for the choice $\epsilon = 0$. The non-zero weights are then
$a=a^\prime=1/2$, while the non-zero matrix elements are $W_1 = W_2 = 1/2$,
which correspond to the switch-and-reverse process illustrated in 
Fig.~\ref{dpaths}(a). This is a deterministic algorithm as the only 
probabilities different from zero are unity.  There is a subtlety 
here as the ratio $c/W_3$ is undetermined for $\epsilon=0$. However, the 
value of this probability can be left undetermined as the vertex with all 
spins down will not be generated as a consequence of the vanishing of the 
weight $W_3$. This is actually more general---whenever a probability cannot 
be defined because of a zero denominator it can be left undetermined 
because the probability of reaching such a vertex is zero in the 
first place. For the XY-model ($\Delta=0$) at zero magnetic field the choice 
$\epsilon=1/4$ gives a different set of zero-bounce rules than the one 
proposed in Ref.~\onlinecite{sse3}, which, however also is a solution of our 
equations (but with $\epsilon=1/2$). It is quite remarkable that for the 
XY-model one can in fact find rules with no bounces for all magnetic field 
strengths up to the saturation field. We expect this to be very useful. 

Outside the shaded region in Fig.~\ref{phasediagram} one or more bounce 
weights must be non-zero. In these regions we will again choose the smallest 
possible values for the bounce weights. Table~\ref{bounceweights} shows 
these values for the different regions in Fig.~\ref{phasediagram}, along with
the minimum value of $\epsilon$ allowed. Selecting a value for $\epsilon$, 
the remaining weights can be obtained using 
Eqs.~(\ref{dim123456}).

\begin{table}
\caption{\label{bounceweights}
Non-zero bounce weights and minimum values of $\epsilon$ for the 
different parameter regions of Solution B of the directed loop equations.
The Roman numerals correspond to those in Fig.~\ref{phasediagram}. We 
have defined $\Delta^\pm = (1 \pm \Delta)/2$.}
\begin{ruledtabular}
\begin{tabular}{llll}
         & bounce weights       &        & $\epsilon_{\rm min}$\\
\rm{I}  &                      &                   & $ (\Delta^--h_b)/2 $ \\
\rm{II} & $b_2 = h_b -\Delta^-$ &$b^\prime_2=-h_b -\Delta^- $ &$0$\\
\rm{III}& $b_2=h_b -\Delta^-$   &                            &$0$\\
\rm{IV} & $b_2=h_b -\Delta^-$   &$b^\prime_3=h_b -\Delta^+$&$0$\\
\rm{V} &                        &$b^\prime_3=h_b -\Delta^+$ &$ (\Delta^--h_b)/2 $ \\
\rm{VI}& $b_3=-h_b-\Delta^+$   &$b^\prime_3=h_b-\Delta^+$ &$-h_b-\Delta/2$\\
\end{tabular}
\end{ruledtabular}
\end{table}

At the boundary between regions in Fig.~\ref{phasediagram}, one of the bounce 
weights vanishes continuously. In 
particular this means that the rules for the Heisenberg antiferromagnet in a 
magnetic field approaches the rules in zero field {\em continuously} as 
$h_b \to 0$. This is to be contrasted to the symmetric Solution A, 
Eqs.~(\ref{symmetricsol}), where the bounce probabilities approach $1/2$ 
as $h_b\to 0$. Hence, the algorithmic discontinuity is indeed removed as
the special deterministic solution at the isotropic point is recovered 
automatically with Solution B (when $\epsilon=0$).

In Sec.~V, the performance of simulations using Solutions A and B will 
be quantified in terms of calculated autocorrelation functions. It will
be shown that the new Solution B can lead to autocorrelation times 
more than an order of magnitude shorter than with Solution A. The
improvements are most dramatic for weak but non-zero fields and weak
Ising anisotropies ($\Delta > 1$). In Sec.~IV we will describe how the 
directed loops also can be adapted to simulations in the path integral 
formalism. Below we first briefly discuss the form of the detailed
balance equations for more general Hamiltonians.

\subsection{General form of the directed loop equations}

The directed loop approach can be easily applied to a much wider class of
models than the $S=1/2$ XXZ-model discussed in the preceding section. The SSE 
operator-loop update with the heat-bath probabilities \cite{sse3} has already 
been applied to several different systems, including spin systems with $S>1/2$
\cite{henelius2}, various boson models \cite{hebert,dorneich1,schmid}, as well 
as the 1D extended Hubbard model \cite{pinaki}. We here briefly outline the 
general form of the directed loop equations and their solutions for a 
general 2-body interaction.

When the operator-loop update is applied to models with higher spins, boson 
or fermion models, it is clear that the simple notion of flipping a spin in 
the $S=1/2$ XXZ-model must be extended to a change in the state at a 
vertex leg where the final state is to one out of several possible ones. 
Consider as an example a spin-$1$ model where a loop can  change the state on 
a leg by one or two units of spin. This is simplified when the total $S^z$ is 
conserved as then these different changes can be considered as two 
{\em independent} loop-updates. This is because changing the state on the exit
leg by two units of spin when the state on the entrance leg is changed by one 
unit, or vice-versa, violates the $S^z$ conservation law. Thus, with such a 
conservation law the state change of the exit leg is uniquely determined given 
the state change at the entrance leg. For simplicity we will here consider 
only those cases where this uniqueness holds, although this is by no means 
a necessary condition. 
    
In order to describe the general form of the directed loop equations for this
type of general two-site interaction it is convenient to change labeling 
somewhat from that used in the previous section. To define this new labeling,
we start by selecting a reference vertex (which can be any of the allowed
vertices) and label its weight $W_1$. We then choose an entrance leg and 
label this leg as leg $1$, and then number the rest of the legs on this 
vertex $2,3$ and $4$. Distributing the weight over all possible exit legs 
according to Eq.~(\ref{sumprob}) gives
\be
        W_1 = a_{11} + a_{12} + a_{13} + a_{14},
\ee
where we have labeled the weights $a_{ij}$ in the extended space by their
entrance ($i$) and exit ($j$) legs.  On changing the states at both the 
entrance and exit legs one arrives at a new vertex. If the entrance and exit 
legs are the same the vertex stays the same. Now label the weight of the 
vertex reached by exiting at leg $i$ as $W_i$. Thus if the exit was on leg 
$2$ we would label that vertex $W_2$. $W_2$ has a similar decomposition 
as $W_1$,
\be
        W_2 = a_{21} + a_{22} + a_{23} + a_{24},
\ee
where now the entrance is on leg two on the vertex which differs from
vertex $1$ by having changed the states at leg $1$ and $2$. The weight 
$a_{21}$ corresponds to the process where the path enters at leg $2$ and 
exits at leg $1$. The states are changed in the {\em opposite} way to that 
when arriving at $W_2$ from $W_1$, and hence the process is undoing the 
changes and we arrive back at $W_1$. From Eq.~(\ref{constraint}) it follows 
that $a_{21} = a_{12}$. Now one can ask the question if exiting at leg 3 or 
4 yields the same vertex type when starting from $W_2$ as it does starting 
from $W_1$. The answer to this is yes, because starting from $W_1$ one would 
change the state at leg $1$ and $3$ while starting from $W_2$ one would 
change the states at legs $2$ and $3$. But $W_2$ differs from $W_1$ only
by having different states at legs $1$ and $2$ and thus the state at leg 
$2$ is {\em changed twice} in opposite directions resulting in the same 
configuration $W_3$. The weights are hence uniquely defined by this
procedure, and one is guaranteed that the only vertices which are related 
by the detailed balance equations are those which can be reached by changing
the state on the entrance leg together with the state on any exit leg of 
the reference vertex. The directed loop equations can therefore be written as
\be
\left(
\begin{array}{cccc}
        a_{11} & a_{12} & a_{13} & a_{14} \\
        a_{12} & a_{22} & a_{23} & a_{24} \\
        a_{13} & a_{23} & a_{33} & a_{34} \\
        a_{14} & a_{24} & a_{34} & a_{44}
\end{array}
\right)
\left(
\begin{array}{c}
        1 \\
        1 \\
        1 \\
        1 
\end{array}
\right)
=
\left(
\begin{array}{c}
        W_1 \\
        W_2 \\
        W_3 \\
        W_4 
\end{array}
\right) 
\ee 
where the matrix on the left hand side is a real symmetric $4 \times 4$ matrix
with all entries non-negative for a useful algorithm. The magnitudes
of the diagonal elements determine the bounce probabilities. This is 
the general 
structure of the directed loop equations for 2-site interactions. There are in
general several such sets of equations, which can be generated one-by-one
by changing the reference vertex and the type of change at the entrance leg.
The reference vertex should then of course be chosen among vertices that
have not yet been generated starting from another reference vertex, in
order not to generate the same equation sets several times.  Some of the 
different sets are typically identical to each other by symmetry, 
as in the case of the $S=1/2$ XXZ model where there are eight sets falling 
into two classes. In that case the structure of the equations changes into 
$3\times 3$ forms because there are only three allowed exit possibilities for 
each entrance leg. To explain this with an example in the scheme used here, 
we can consider the vertex with all spins down as the reference vertex. Then 
$W_2 =0$ as this configuration corresponds to the case where the lower legs 
($1$ and $2$) are flipped, resulting in a vertex with weight zero. This 
immediately implies that all $a$'s (being all non-negative) with an index $2$ 
must be zero and so the result is that row $2$ and column $2$ is taken out 
resulting in a $3 \times 3$-matrix. In general, there can be a large numbers 
of $4 \times 4$ equation sets, some or all of which reduce into $3\times 3$ 
and $2\times 2$ sets (e.g., for Hubbard-type electron models there are
both $2\times 2$ and $3\times 3$ sets, but no $4\times 4$ sets).

Let us consider the $3 \times 3$ case in greater detail and ask when one can 
do without bounces, as we saw was possible in a region in parameter space of
the $S=1/2$ XXZ-model discussed in the previous section. To do this, it
is convenient to first label the equations so that $W_3 \ge W_2 \ge W_1$. 
We then set all the diagonal entries (the bounce weights) to zero and find the
region of different $W$'s for which the equation set has strictly positive 
solutions. In this case the solution is unique as there are 3 equations and 
3 unknowns and it is easy to see that the $a$'s are positive only when 
$W_3 \le W_1 + W_2$, and hence one find a directed loop solution without 
bounces only when this condition is satisfied.

Allowing bounces, it is also easy to see that one can always do with only one
bounce, the one which bounces off of the vertex with the largest weight.
If $W_3$ is the largest weight one can set $a_{11}=a_{22}=0$ and 
$a_{33} = W_3-W_1-W_2$, which gives $a_{12}=0$, $a_{13}=W_1$ and $a_{23}=W_2$.
This means that the probability for moving between the configurations with
smallest weight is zero while that of moving from the largest weight
configuration to the smaller ones is the ratio of the smaller weight to the
larger weight and unity for the reverse process. The bounce probability is
unity minus the probability for moving to the smaller weight configurations.
A similar analysis can be carried out for the $4\times 4$ equation sets
appearing for $S>1/2$ models and boson models.

The equation sets involving larger matrices, as encountered when dealing with
interactions involving more than two sites, can also be studied in a similar 
manner. It should be pointed out that there is nothing that guarantees
{\it a priori} that the operator-loop update is ergodic (in combination
with the diagonal updates), for any solution of the directed loop equations. 
Ergodicity requires that all allowed vertices can be generated through a 
series of loop updates, and this is typically the case with 2-particle terms 
(although one could in principle construct models where it is not the case). 
However, simple one-dimensional loops such as those discussed here cannot 
always accomplish this alone when the interaction includes more than two 
particles, even in the case of relatively simple models. The SSE method has 
recently been applied to an $XY$-model with a standard $2$-spin interaction 
$J$ and a $4$-spin term $K$ \cite{jkmodel}. In that particular case, an 
operator-loop update can be used, and is ergodic, for $|J| > 0$, but for $J=0$ 
another cluster-type update had to be carried out. In practice, a combination
of the two updates had to be used for large $K/J$.

\section{Path integral formulation \label{PIMsection}}

In this section we will discuss how the directed loops can be applied to 
the path integral Monte Carlo method (PIM) formulated in imaginary time. 
Such methods are known as world-line methods in discrete \cite{worldline} 
or continuous \cite{prokofev,beard} imaginary time. The close relationships 
between the SSE and PIM representations of quantum statistical mechanics 
have been explored in previous works \cite{irsse,athens}. Here we will show 
that also the directed loop ideas can be almost directly translated 
from SSE into the PIM formalism.

\subsection{Construction of the path integral}

\begin{figure}
\includegraphics[clip,width=5cm]{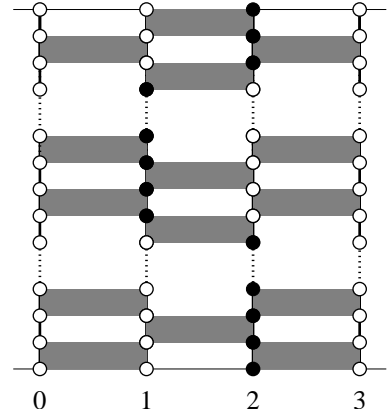}
\caption{The checkerboard breakup of the space-time for a spin chain with four
sites with open boundary conditions. $H_1$ have terms acting  on the links 
between site $0$ and $1$ and the link between site $2$ and $3$. $H_2$ acts 
on the link between site $1$ and $2$. The shaded plaquettes show where the 
Hamiltonian acts.
\label{Checkerboard}} 
\end{figure}

We start by writing the partition function as
\be
   Z = {\rm Tr} \lbrace e^{-\beta H} \rbrace
     = {\rm Tr} \left\lbrace \prod_{t=1}^L e^{-\Deltatau H}\right\rbrace, 
\ee
where $\Deltatau = \beta/L$ and $L$ is a large integer. The Hamiltonian is 
generally a sum of non-commuting pieces, and in order to deal with the 
exponential it is convenient to employ the Suzuki-Trotter trick 
\cite{suzuki1}. This involves dividing the Hamiltonian into several sets 
of terms, where all terms within a set are commuting while the sets themselves
are non-commuting. Because the Hamiltonian is multiplied by the small quantity
$\Deltatau$ it is possible to split the exponential into a product of 
exponentials, each having one set in the exponent. The errors arising from
this approximation vanishes as $\Deltatau \to 0$ \cite{suzuki1,fye}. Consider 
as an example the XXZ chain. Then the Hamiltonian can be divided into
two sets, one involving the operators which act on sites $2n$ and $2n+1$ while
the other set involves the operators acting on sites $2n+1$ and $2n+2$. 
It is then possible to insert complete sets of states, which can be chosen 
to be written in terms of $S^z$-components, between all the exponentials and 
the partition function can be written \cite{suzuki2,barma,worldline,cullen}
\be
	Z = \sum_{\left\{\sigma \right\}} 
 \prod_{t=1}^L \langle \sigma_{t+1}| e^{-\Deltatau H_2} |\sigma_{t+1/2} \rangle
	   \langle \sigma_{t+1/2}| e^{-\Deltatau H_1} |\sigma_{t} \rangle ,
\ee
where $\sigma$ is a shorthand for a spin configuration in the $S^z$-basis 
of all sites in the chain. The sum is over all possible sets of spin
configurations, two complete sets of states for each time step $t$, 
and the trace implies 
$\sigma_{L+1}=\sigma_{1}$. This is called the checkerboard breakup as one 
can visualize it as a checkerboard pattern (see Fig.~\ref{Checkerboard}) 
where all the matrix elements are pictured as shaded plaquettes. 
This breakup is completely general and can also be used for higher-dimensional
lattices. Because each set $H_1$ and $H_2$ consists of individually commuting 
terms it suffices to consider the interaction on one shaded plaquette only 
and the matrix elements can easily be written down. Keeping only terms to 
first order in $\Deltatau$ one finds 
\bea
	W_1 & = & \langle \up \down | e^{-\Deltatau H} |\down \up \rangle
	         =\langle \down \up | e^{-\Deltatau H} | \up \down \rangle 
                 = \Deltatau /2 ,
	\nonumber \\
	W_2 & = & \langle \up \down |e^{-\Deltatau H}| \up \down \rangle
	         =\langle \down \up |e^{-\Deltatau H}| \down \up \rangle 
	\nonumber \\
	& = & 1+ (C+\Delta/4) \label{ME} \Deltatau ,\label{pimw1234}\\
	W_3 & = & \langle \down \down |e^{-\Deltatau H}| \down \down \rangle 
	         = 1+ (C-\Delta/4-h_b) \Deltatau , \nonumber \\
	W_4 & = & \langle \up \up |e^{-\Deltatau H}| \up \up \rangle 
	         = 1+ (C-\Delta/4+h_b) \Deltatau . \nonumber
\eea
These matrix elements differ from the matrix elements (\ref{matrelem})
in the SSE method only in that the Hamiltonian is multiplied by the factor 
$\Deltatau$ and the diagonal matrix elements also come with the zeroth-order 
term of the exponential. The weight $W_1$ comes with a minus sign 
which here is omitted by implicitly performing a $\pi$-rotation about the 
$S^z$-axis for spins on one sub-lattice. This can be done whenever the 
lattice is bipartite. One can of course also calculate the matrix elements 
(\ref{pimw1234}) exactly, but since we will here take the continuum limit it
is sufficient to go to linear order in $\Delta\tau$, where the similarity
to the SSE expressions are most evident.

In the ordinary world-line loop algorithm (for a review see 
Ref.~\cite{evertzchapter}), two loop segments are assigned to each and every 
shaded plaquette in a stochastic way. The shaded plaquettes are corner-sharing
so that when all shaded plaquettes have been assigned segments one can identify closed loops.
Given that the probabilistic rules for the assignment of loop segments for 
each shaded plaquette follows the analogy of Eq.~(\ref{constraint}) and 
Eq.~(\ref{sumprob}), one can flip a loop with any probability. In particular
one can pick a random site and a random imaginary time and flip the loop 
which includes this point with probability unity. One can also turn this 
around and {\em first}, before any loop is constructed, pick a random point 
in space-time and then construct the loop starting at this point and flipping 
spins with probability unity as the loop is being constructed.

\begin{figure}
\includegraphics[clip,width=3.5cm]{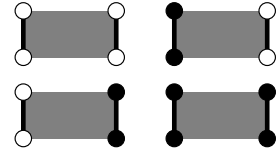}
\caption{Loop and spin configurations which should have the same weight when 
allowing the loops to be flipped independently.
\label{PIMconstraints}}
\end{figure}

When assigning loop segments to each shaded plaquette one needs two
loop segments for each plaquette in order to fill the lattice completely.
Then many configurations can be reached, as one should be able
to independently flip spins along one or both the loop segments. Thus
one gets relatively many constraints of the type (\ref{constraint}). This is 
illustrated in Fig.~\ref{PIMconstraints}. In fact, in zero field there are 
just as many equations as unknowns, and this set has only non-negative 
solutions in the XY-like case $-1 \leq \Delta \leq 1$. In a magnetic field 
there is one additional equation and the set does not have any solutions. 
Within the standard loop algorithm this is repaired by introducing additional 
processes which ``freeze'' loops together, i.e., if spins on one loop is 
flipped, spins on any loops frozen together with the first one will also be 
flipped. This increases the number of unknowns in the equation set, making a 
solution possible. While we are not aware of any systematic studies of the 
effects of the freezing process, it tends to freeze all loops together 
resulting in the trivial spin update where all spins are flipped. It is 
therefore not very effective. However, in the extreme Ising limit the 
freezing is responsible for the fact that the loop algorithm becomes
equivalent to the Swendsen-Wang algorithm, 
and hence the freezing of loops has some merits.

Another method to make the loop algorithm work in a magnetic field is to 
apply the field in the $x$-direction, thereby changing the matrix 
elements and introducing a minus sign. Using the concept of merons the
resulting sign problem can be solved \cite{chandrasekharan,cox}, but the
simulation algorithm is not very efficient for large systems. If one relaxes 
the condition that the loops should be flipped with probability one and 
instead chooses weights such that the flipping probability is maximized, it 
is possible to find rules that work very well at extreme fields 
\cite{syljuasen}. However, this success at extreme fields must be regarded 
as a lucky circumstance and is not generally valid for lower fields. Yet 
another and perhaps the simplest loop method in the presence of a magnetic 
field is to construct the loops as if the field was absent and then include 
a Metropolis decision whenever attempting to flip a loop that changes the 
magnetization. This method is, however, very ineffective \cite{dorneich2} 
(except at extremely weak fields; $h/J \alt 1/(\beta N)$ \cite{troyer}) as 
is to be expected as it does not take into account the actual physics of 
the model which is the competition between the magnetic field and the 
exchange energy. 

None of the above methods for treating external fields has proven as useful 
in practice as the SSE operator-loop algorithm \cite{sse3}. The worm algorithm
for path integral simulations in continuous imaginary time \cite{prokofev}
shares some important features with the SSE operator-loops (specifically, 
there is an analogue to the back-tracking) and has also been used successfully.
However, its autocorrelation times appear to be longer (as can be seen in 
comparing our results in Sec.~V with those presented in 
Ref.~\onlinecite{kashurnikov}). We will discuss differences between the 
procedures used to construct directed loops and worms in Sec.~VII. Because 
the directed loops is a further improvement of the SSE approach, it is natural
to investigate if these concepts can also be implemented in the path 
integral formulation. 

\subsection{Directed loops in the PIM}

To implement the notion of directed loops in the path integral formulation 
we note the similarities of the vertices in the SSE and the shaded plaquettes
in the PIM. We can identify a corner of a shaded plaquette with a vertex leg 
in the SSE. Both have a spin attached, and each corner (leg) is connected to 
another corner (leg) on another shaded plaquette (vertex). To construct a
directed loop, we first choose a 
random entrance corner at a random shaded plaquette. Then, depending on the 
spin configuration, we choose an exit corner and place a directed loop segment
between the entrance corner and the exit corner. The spins connected by the 
loop segment are flipped with probability one. The spin on the 
exit corner is then the entrance spin of the next shaded plaquette and the 
process continues until the loop closes. In contrast to the usual loop 
algorithm there is no notion of freezing loops, but there is the necessary 
(at least in some regimes) process of bouncing where the ``loop head'' 
backtracks some distance along its path and reverses spin flips.
  
Because of the relation between the SSE vertices and the shaded plaquettes, 
and the similarity of the matrix elements (\ref{w1234}) and (\ref{pimw1234}),
one can immediately write down the the detailed balance equations for the PIM
using Fig.~\ref{all} and interpreting the vertices as shaded plaquettes. 
As in the SSE, there are eight sets of directed loop equations which are 
reduced to two by symmetries. Substituting the plaquette weights and expressing
the extended configuration weights in terms of the bounce weights we get 
\bea
	a & = & \left( \f{1+\Delta}{4} + \f{h_b}{2} \right) \Deltatau 
	+ \f{-b_1-b_2+b_3}{2},
	\nonumber \\
	b & = & \left( \f{1-\Delta}{4} - \f{h_b}{2} \right) \Deltatau + 
	\f{-b_1+b_2-b_3}{2},
	\nonumber \\
	c & = & 1+ \left(C -\f{1}{4} - \f{h_b}{2}\right) \Deltatau 
	+ \f{b_1-b_2-b_3}{2},
	\nonumber \\
	a^\prime & = & \left( \f{1+\Delta}{4} - \f{h_b}{2} \right) \Deltatau 
	+ \f{-b^\prime_1-b^\prime_2+b^\prime_3}{2},
	\label{PIMeqs} \\
	b^\prime & = & \left( \f{1-\Delta}{4} + \f{h_b}{2} \right) \Deltatau + 
	\f{-b^\prime_1+b^\prime_2-b^\prime_3}{2},
 	\nonumber \\
	c^\prime & = & 1+ \left(C -\f{1}{4} + \f{h_b}{2}\right) \Deltatau 
	+ \f{b^\prime_1-b^\prime_2-b^\prime_3}{2}.
	\nonumber
\eea 
Non-negative weights are required to avoid sign problems. This
implies that there are regions where bounces must be non-zero. In fact
the same ``algorithmic phasediagram'' as shown in Fig.~\ref{phasediagram} 
applies here, with the exception that in this case there are no
restrictions on $C$ (or $\epsilon = C-\Delta/4-h_b$) as 
it always occurs multiplied by $\Deltatau$ in
a combination where there also is the zeroth-order term of the exponential. 
In fact, in the construction of the loops in continuous imaginary time, where
only quantities to order $\Deltatau$ matters, the value of $C$ drops out 
completely as we will consider {\em ratios} where it turns out that $C$ 
does not occur to 
order $\Deltatau$. Thus in contrast to the SSE, there is nothing gained by 
adjusting $C$ in the path integral representation. Whenever 
in a region of parameter space where bounces are needed, one can choose them 
to be the minimum values as summarized in Table \ref{bounceweights}, with 
the only modification that the bounce weights should be multiplied by 
$\Deltatau$. As in the SSE method the actual probability for choosing an 
exit corner, given an entrance corner and a spin configuration on a shaded 
plaquette, is obtained by dividing one of the weights above by the 
appropriate matrix element from Eqs.~(\ref{ME}).

In the limit $\Deltatau \to 0$ this method might seem very slow as one needs
to make a choice for every plaquette of which there are infinitely many in 
this limit. However, one can use the method employed in the continuous time 
implementation of the standard world-line
loop algorithm \cite {beard}, which is based on the fact 
that the $c$,$c^\prime$ weights are of order unity. The $c$,$c^\prime$ weights
describe the process of continuing the loop construction in the imaginary 
time direction on the same site. Being of order unity means that this will be 
the dominating process. The other processes are multiplied by $\Deltatau$ and 
will therefore occur much less frequently. 

\begin{figure}
\includegraphics[clip,width=7.5cm]{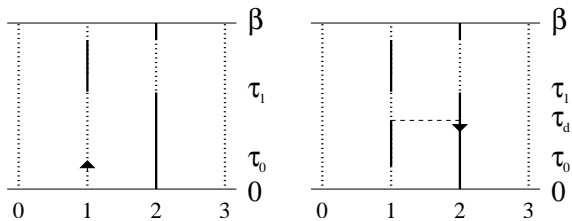}
\caption{Left: Continuous imaginary time construction of a loop. This figure 
can be understood as the limit $\Deltatau \to 0$ of Fig.~\ref{Checkerboard}, 
dotted (solid) lines correspond to spin down (up). Starting at an arbitrary 
site and time (indicated by the arrow) 
a probability of ``decay'' dependent on the spin 
states of the neighbors is calculated, and the loop head is moved to the 
point of decay. Right: A resulting $a^\prime$ decay at a time $\tau_d$ 
where the segment up to the decay has changed orientation and a new arrow 
is placed.}
\label{example}
\end{figure} 

To illustrate in detail how a loop is constructed in the limit $\Deltatau 
\to 0$, consider as an example the situation shown in Fig.~\ref{example}. 
This figure shows the full imaginary-time spin configurations for four sites. 
The dotted (solid) lines correspond to spin down (up). The figure can be 
understood as the limit $\Deltatau \to 0$ of Fig.~\ref{Checkerboard}.  
The loop construction consists of moving the loop head. This motion begins 
at a random site and time in a random direction. In Fig.~\ref{example} the 
starting point and direction is marked by an arrow. From the arrow at time 
$\tau_0$ to the time $\tau_1$ the spin configuration on site $1$ and its 
neighbors $0$ and $2$ stay unchanged. At time $\tau_1$ there is a spin-flip 
process exchanging the spins on sites $1$ and $2$. This means that half of 
the $2\tau /\Deltatau$, $\tau=\tau_1-\tau_0$, shaded plaquettes (the factor 
$2$ is from the fact that there are two neighbors) between the starting point 
$\tau_0$ and $\tau_1$ are of the type $W_2$, while the other half is of the 
type $W_3$. The loop head will therefore enter alternately the lower left 
corner on shaded plaquettes having weight $W_2$ and the lower right corner 
on shaded $W_3$ plaquettes. On exiting the shaded $W_2$ plaquette, one of 
the three processes $a^\prime$, $b^\prime_2$ or $c^\prime$ can happen, 
while for each of the $W_3$ plaquettes one of the processes $b$,$c$ or $b_3$ 
can happen. The $c$ and $c^\prime$ processes are by far the most probable 
as they are of order unity while the others are of order $\Deltatau$. 
Therefore until one of the other processes of order $\Deltatau$ occurs, 
the loop head will just continue its motion in the upward direction on 
site $1$. The probability for the first occurrence of one of the processes 
of order $\Deltatau$ within an interval $\Deltatau$ after time $\tau^\prime$ 
is given by
\bea
	P \left( \tau^\prime \right) \Deltatau 
	& = & \left( \f{c^\prime}{W_2} \f{c}{W_3}
	      \right)^{\tau^\prime/\Deltatau}
	  	\left( 1 - \f{c^\prime}{W_2}
	  	     + 1 - \f{c}{W_3} \right) \nonumber \\
	& = & e^{-(\alpha_{0} + \alpha_{2}) \tau^\prime} 
              (\alpha_{0}+\alpha_{2}) \Deltatau, \label{decayprob}
\eea
where in the last equality we have taken the limit $\Deltatau \to 0$, and
the quantities $\alpha_i$ are finite as $\Deltatau \to 0$;
\bea
	\alpha_{0} & = & \f{b+b^\prime_3}{W_3 \Deltatau}, \\
	\alpha_{2} & = & \f{a^\prime+b^\prime_2}{W_2 \Deltatau},
\eea
where the subscript on $\alpha$ indicates which neighbor is considered.
Recall that by definition $W_3=b+c+b^\prime_3$ and $W_2=a+b_2+c$. 
Thus, with a random number generator one can generate ``decay'' times 
according to the distribution (\ref{decayprob})  and take the random decay 
time generated as the point where one of the processes $a^\prime$, 
$b^\prime_2$, $b$ or $b_3$ occurs. If the decay time so generated is bigger 
than $\tau_1-\tau_0$ the loop head can be moved directly all the way 
to time $\tau_1$, while flipping all the spins on site $1$ up to time 
$\tau_1$. There it enters a shaded plaquette from the lower left corner. 
This plaquette has weight $W_1$, and the possible choices for exit corners 
are determined by the ratio of the weights $b^\prime_1$,$a^\prime$ and 
$b^\prime$ to $W_1$ which all are finite as $\Deltatau \to 0$. One can hence 
just use the random number generator to select the exit corner. Given that the 
outcome of this choice is, for instance, $a^\prime$ 
the loop head would move to site $2$ while flipping 
spins which changes the shaded plaquette of type $W_1$ to be of type $W_2$. 
The process would then continue in the downward direction on site $2$.  
If the decay happens before $\tau_1$ the loop head moves to the decay point 
while flipping spins and then a choice between the possible decay types is 
made. Given that a decay occurs, the choice of different types of decays is 
again independent of $\Deltatau$ as only the ratios matter. As an example, 
the probability of selecting $a^\prime$ is 
$a^\prime/(a^\prime+b^\prime_2+b+b_3)$. This type of process is illustrated in
Fig.~\ref{example}. Having made the choice the process 
continues, and the loop closes when the loop head reaches the original 
starting point. 

In practice it is convenient to store the spin-flip events in a doubly-linked
list for each lattice site so that spin-flips can be added and removed 
efficiently. The main computational cost is then to search the site of
the loop head and its neighbors for spin transitions. 
   
In zero magnetic field the directed PIM loop algorithm proposed here 
corresponds exactly to the single-cluster formulation of the ordinary loop 
algorithm for $-1 \leq \Delta \leq 1$ \cite{beard,evertz}. This can be seen 
by setting all bounce weights to zero and $C=-\Delta/4$, and then comparing 
our weights to Eq.~(39) in Ref.~\cite{evertzchapter}. In the language of the 
usual loop algorithm, our weight $a$ corresponds to horizontal breakups, $b$ 
to diagonal breakups, and $c$ to vertical breakups. The general algorithm with
bounces is more similar to the worm algorithm \cite{prokofev}, but the
processes by which the worm is propagated through space-time are different
and do not correspond to a solution of our directed loop equations. This
will be further discussed in Sec.~VII. In Sec.~V we will demonstrate that the 
directed loop processes, especially with Solution B (in both SSE and PIM
implementations) lead to much more efficient simulation algorithms.

\section{Autocorrelations}

Autocorrelation functions provide quantitative measures of the efficiency of
a Monte Carlo sampling scheme in generating statistically independent
configurations. For a quantity $Q$, the normalized autocorrelation function is
defined as
\begin{equation}
A_Q(t) = {\langle Q(i+t)Q(i) \rangle - \langle Q(i) \rangle ^2 \over 
\langle Q(i)^2 \rangle - \langle Q(i) \rangle^2 },
\label{atau}
\end{equation}
where $i$ and $t$ are Monte Carlo times, for which we will use the unit of
1 MCS (as defined in Sec.~II-D in the case of SSE, and with an analogous
definition for the PIM). The brackets indicate the average over the 
time $i$. Asymptotically, the autocorrelation function decays exponentially 
as $\sim {\rm e}^{-t/\tau_Q}$, where the asymptotic 
autocorrelation time $\tau_Q$ is given by the slowest mode of the simulation
(the transition matrix of the Markov chain) to which the observable $Q$ 
couples. For short times, the behavior is typically different for 
different quantities, even if $\tau_Q$ is the same. The integrated 
autocorrelation time is defined according to 
\begin{equation}
\tau_{\rm int}[Q] = {1\over 2} + \sum_{t=1}^\infty A_Q(t)
\end{equation}
and is the autocorrelation measure
of greatest practical utility \cite{evertzchapter}. 

In this section we will present integrated autocorrelation times for some 
important quantities in several regions of the parameter space of the 
anisotropic Heisenberg model (\ref{ham1}). We cannot present a completely 
exhaustive study, however, since in addition to the field $h$ and the
anisotropy $\Delta$, the autocorrelations also depend on temperature 
$T/J = \beta^{-1}$ and the lattice size. In addition, in SSE simulations the 
autocorrelations depend on the constant 
$\epsilon$ in the matrix elements (\ref{matrelem}).
One of our aims here is to find the optimum value if $\epsilon$. We compare 
simulations with the original general (non-deterministic) SSE operator-loop 
update \cite{sse3} (Solution A) and the new solution of the directed loop 
equations discussed in Sec.~III B (Solution B). We also present some results 
obtained with Solution B in continuous-time PIM simulations.

The physical quantities that we will focus on here are the magnetization, 
\begin{equation}
M = {1\over N} \sum_{i=1}^N \langle S^z_i \rangle,
\end{equation}
the uniform magnetic susceptibility
\begin{equation}
\chi_u = {\beta\over N} \left \langle \left ( \sum_{i=1}^N S^z_i \right )^2
\right \rangle ,
\end{equation}
the staggered susceptibility,
\begin{equation}
\chi_s = {1\over N} \sum\limits_{k,l}(-1)^{x_k-x_l+y_k-y_l}
\int\limits_0^\beta d\tau \langle S^z_k (\tau)S^z_l (0) \rangle ,
\label{chisdef}
\end{equation}
and the spin stiffness, 
\begin{equation}
\rho_s = {\partial^2 E(\phi)\over \partial\phi ^2},
\end{equation}
where $E(\phi)$ is the internal energy per spin in the presence of
a twist $\phi$ in the boundary condition. These quantities and their SSE
estimators have been discussed in detail in Ref.~\cite{sse4}.

We note again that the definition of an MCS in the generic SSE operator-loop 
scheme involves some degree of arbitrariness, as was discussed in Sec.~II D. 
There is also a statistical uncertainty due to the statistical determination 
of the number $N_l$ of operator-loops constructed per MCS. In all the SSE
simulations discussed here, $N_l$ was adjusted during the equilibration of 
the simulation so that on average $2M$ vertex legs (excluding bounces)
were visited in each MCS. The maximum expansion power $M$ was increased if 
needed after each equilibration MCS, so that $M = 1.25 \times n_{\rm max}$, 
where $n_{\rm max}$ is the highest power $n$ generated so far in the 
simulation. The statistical uncertainties in $N_l$ and $n_{\rm max}$ imply 
some fluctuations in the definition of an MCS. This, in turn, results in 
fluctuations in the results for the integrated autocorrelation times that can 
be larger than their statistical errors. Typically, these fluctuations are 
only a few percent, however, and are hence not problematic. 

In the PIM simulations, we adjusted $N_l$ so that on average the total length 
(again excluding bounces) of all $N_l$ loops in a MCS is equal to $\beta N$; 
the space-time volume. The definitions of an MCS in SSE and PIM simulations 
are hence similar but not identical. One reason why it is difficult to
construct exactly comparable MCS definitions in the SSE and the PIM is that 
the diagonal single-operator updates carried out separately in SSE are in 
effect accomplished during the loop construction in the PIM. Another 
difference is that there is no adjustable constant $\epsilon$ in the PIM. 
In Ref.~\cite{dorneich2} an alternative approach of normalizing the 
autocorrelation times by the actual number of operations performed was 
used. However, also this definition may be ambiguous since it depends on 
the details of the implementation, and there are also differences in the 
actual CPU time consumed depending on the mix of operations (integer, 
floating point, boolean, etc.). These issues are not of major significance 
in the calculations we present below, but should nevertheless be kept in 
mind when comparing autocorrelations for the two methods. 

The reminder of this section is organized as follows: In A we first discuss 
SSE simulations of the 1D Heisenberg model in an external field. In B we
consider SSE simulations of 2D systems in fields and with anisotropies. PIM
results for both 1D and 2D systems are presented in C. We have also studied 
several isotropic systems at critical points and extracted the dynamic 
exponent of the simulations. We discuss these results in D.

\begin{figure}
\includegraphics[clip,width=7.5cm]{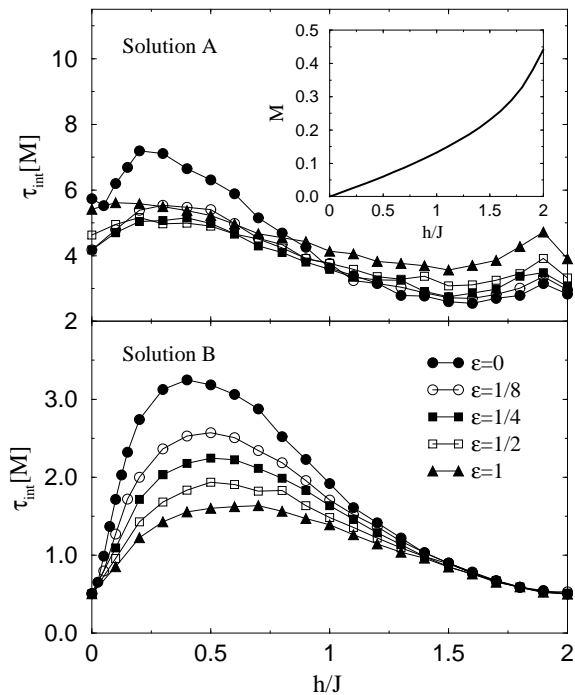}
\caption{Integrated autocorrelation times vs external field for the 
magnetization of an $N=64$ Heisenberg chain at $\beta=16$. The upper
and lower panels show results of simulations using Solutions A and B,
respectively. Several values of the constant $\epsilon$ were used, as 
indicated by the legends in the lower panel. The inset shows the 
magnetization itself.} 
\label{l64_m}
\end{figure}

\subsection{SSE simulations in 1D}

When the constant $\epsilon=0$, the vertices with all spins up or all spins 
down are excluded from the SSE configuration space when $h=0$, since
the corresponding matrix elements (\ref{matrelem}) then vanish. When $h > 0$, 
the all-up vertex is again allowed. With $\epsilon > 0$ all vertices are
allowed and the propagation of the loop is then more random. We here begin 
by studying how the simulation efficiency depends on $\epsilon$ in the case
of the 1D Heisenberg model ($\Delta=1$) in a field.

\begin{figure}
\includegraphics[clip,width=7.5cm]{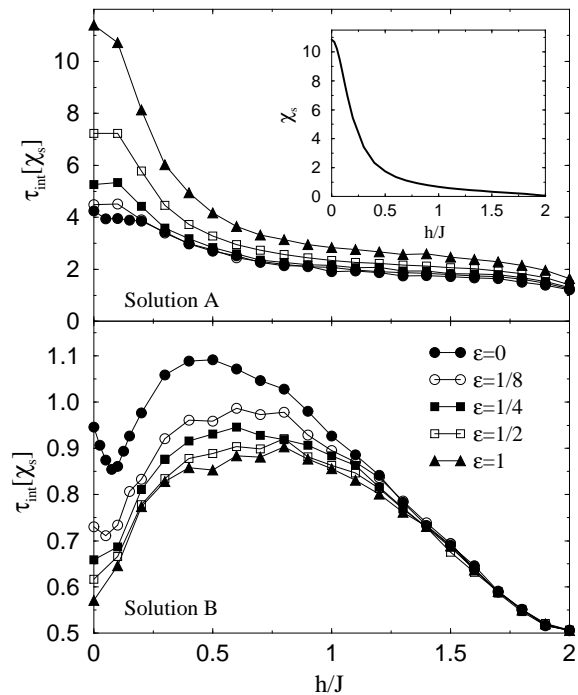}
\caption{Integrated autocorrelation times vs external field for the 
staggered susceptibility of an $N=64$ Heisenberg chain at $\beta=16$.
The inset shows the staggered susceptibility.}
\label{l64_x}
\end{figure}

Figs.~\ref{l64_m} and \ref{l64_x} show the field dependence of the integrated 
autocorrelation time of the magnetization and the staggered susceptibility in 
simulations of chains with $64$ sites at inverse temperature $\beta=16$. As 
shown in the inset of Fig.~\ref{l64_m}, at the $T=0$ saturation field 
($h_{\rm sat}/J = 2$ in 1D) the magnetization is about 10\% from saturation 
at this temperature. The staggered susceptibility is peaked at $h=0$, 
reflecting the fact that the staggered spin-spin correlation function 
for spin components parallel to the external field is 
dominant only in the absence of a field.
In the case of Solution A simulations, the effect of increasing $\epsilon$ 
from $0$ is an initial small drop in $\tau_{\rm int}[M]$ for fields 
$h \alt 0.8$ and a small increase at higher fields. There is a substantial 
increase in $\tau_{\rm int}[\chi_s]$ for weak fields. As $\epsilon$ is further
increased there is a small increase in $\tau_{\rm int}[M]$ also for weak
fields. In contrast, with Solution $B$ increasing $\epsilon$ has favorable 
effects on both autocorrelation times up to the highest $\epsilon$ studied 
here. The effects are very small for high fields, however, since there the 
autocorrelation time is already close to its lower bound $0.5$ when 
$\epsilon=0$. For all $\epsilon$-values, the autocorrelation times are 
considerably shorter with Solution B than with Solution A. This shows that the 
strategy of decreasing the probability of the bounce processes in the 
operator-loop construction is working. The effects are particularly pronounced
at and close to $h = 0$, where the shortest autocorrelation times with 
Solution B are only about $10\%$ of those with Solution A. 

\begin{figure}
\includegraphics[clip,width=7.5cm]{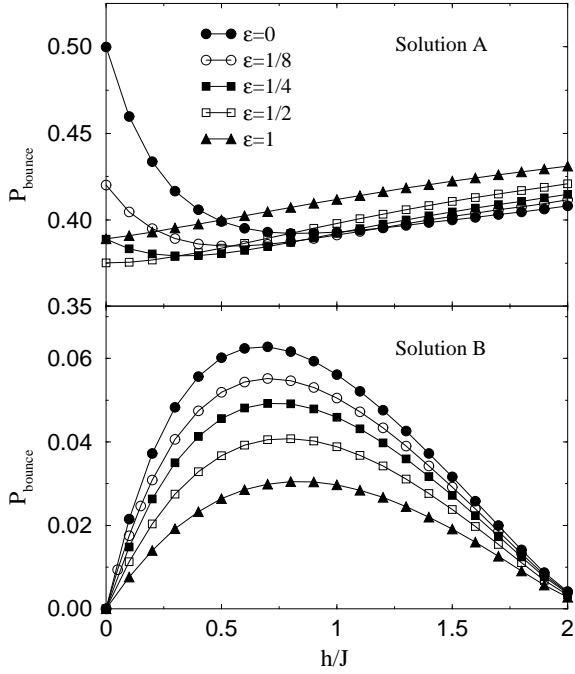}
\caption{Bounce probabilities in Solution A and B simulations of an 
$N=64$ Heisenberg chains at $\beta=16$, using different values 
of $\epsilon$.}
\label{bounce}
\end{figure}

In Fig.~\ref{bounce} we show the probability of bounces in the simulations 
($P_{\rm bounce}$ is the fraction of bounces, including length-$0$ 
loops). The behavior reflects that of the autocorrelation times. With Solution
B, $P_{\rm bounce}$ decreases monotonically with $\epsilon$ for all fields,
whereas with Solution A the behavior is non-monotonic. In Solution B, the 
vanishing of $P_{\rm bounce}$  both in the limits $h \to 0$ and 
$h \to h_{\rm sat}/J$ (at $T=0$) follows by construction, as discussed in
Sec.~II. With Solution A the bounce rate is large in these limits.

\begin{figure}
\includegraphics[clip,width=7.5cm]{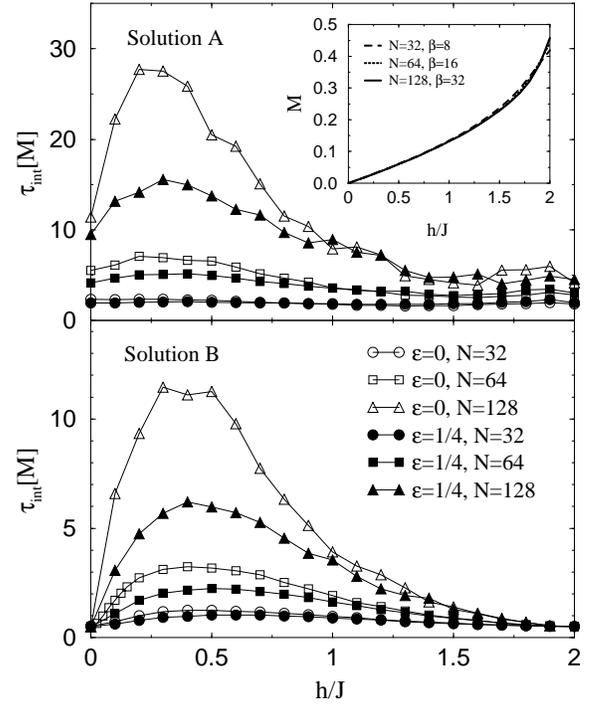}
\caption{Integrated autocorrelation times for the magnetization in 
simulations of chains of different lengths $N$ at inverse temperature 
$\beta=N/4$. The inset shows the magnetization.}
\label{mltau}
\end{figure}

For weak fields, a small $\epsilon > 0$ has favorable effects on the 
magnetization autocorrelations both with solution A and B. In the case of
Solution B, both $\tau_{\rm int}[M]$ and $\tau_{\rm int}[\chi_s]$ continue 
to decrease also when $\epsilon \approx 1$, as seen in Figs.~\ref{l64_m} and 
\ref{l64_x}. Nevertheless, it is not practical to use a very large $\epsilon$ 
since the average expansion order $\langle n\rangle$ (and hence the operator 
sequence size $M$) has a contribution $\epsilon \beta N_b$, and there is
a similar increase in the number of operations needed to carry 
out one MCS. However, Figs.~\ref{l64_m} and \ref{l64_x} indicate that even 
a small value, ($\epsilon \sim 1/4$), gives a significant improvement 
of the magnetization autocorrelations relative to $\epsilon = 0$ simulations. 
We find that this behavior persists also for larger system sizes and lower 
temperatures. Fig.~\ref{mltau} shows $\tau_{\rm int}[M]$ for different 
system sizes $N$ at inverse temperature $\beta=N/4$, using both $\epsilon=0$ 
and $1/4$. The advantage of $\epsilon=1/4$ becomes more pronounced with 
increasing system size. For $N=128$ the maximum $\tau_{\rm int}[M]$ is reduced
by about $50\%$ for both Solution A and B. The relative advantage of 
Solution B over A is again the most dramatic in the limit $h \to 0$.
In both solutions, the autocorrelation time is rather strongly peaked, with
the peak position for the largest systems at slightly higher fields for
Solution B. The reason for this type of field dependence is not clear and 
deserves further study. It cannot be ruled out that a still more efficient 
directed loop solution could be found at intermediate field strengths (which 
would imply that minimizing the bounce probability does not necessarily 
lead to the most efficient algorithm). 

\begin{figure}
\includegraphics[clip,width=7.5cm]{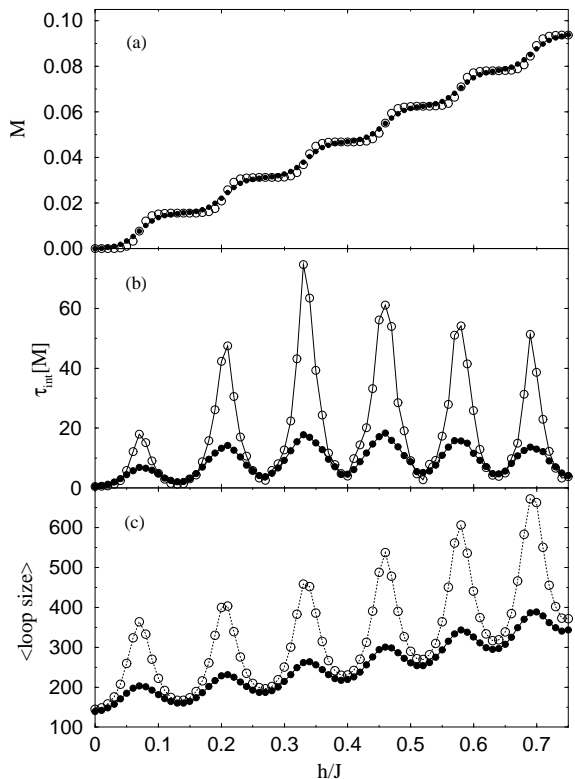}
\caption{Magnetization vs field of an $N=64$ chain (a), the corresponding 
integrated autocorrelation time (b), and the average size of the 
operator-loops (c). Solid and open circles show results at $\beta=64$ and 
$128$, respectively. The simulations were carried out with Solution B of 
the directed loop equation with $\epsilon=1/4$.}
\label{steps}
\end{figure}

When the temperature becomes small compared to the finite-size gaps in
the system, a step structure in the magnetization versus field curve
can be clearly resolved, as is shown in Fig.~\ref{steps}(a). These steps are
also reflected in the autocorrelation time, as shown in Fig.~\ref{steps}(b). 
There are sharp maximas in the regions where the magnetization switches between
two values. Exactly at $T=0$, the autocorrelation function (\ref{atau}) for 
the magnetization is ill-defined, since there are then no fluctuations in
$M$ on the magnetization plateaus. However, we find that the limit $T\to 0$ is
well-behaved in the simulations. Exactly at the switching fields, 
$\tau_{\rm int}[M]$ appears to diverge, however, showing that tunneling 
between the two equal-probability magnetization sectors becomes 
rare. Fig.~\ref{steps}(c) shows the average size of the operator-loops. 
There are maxima at the switching fields, with the peak heights growing as
the temperature is lowered. On the plateaus, the loop size does not change
much with $\beta$. A divergence of the average loop size with $\beta$ at the 
switching fields can be expected, since in order for the magnetization to 
change, the loop has to wrap around the system in the SSE propagation (or 
imaginary time) direction, which is of length $\sim \beta$. The convergence 
of the average loop size on the plateaus can be understood on the same 
grounds. Apart from the oscillations, there is also a significant increase 
in the loop size as the field increases.

\begin{figure}
\includegraphics[clip,width=7.5cm]{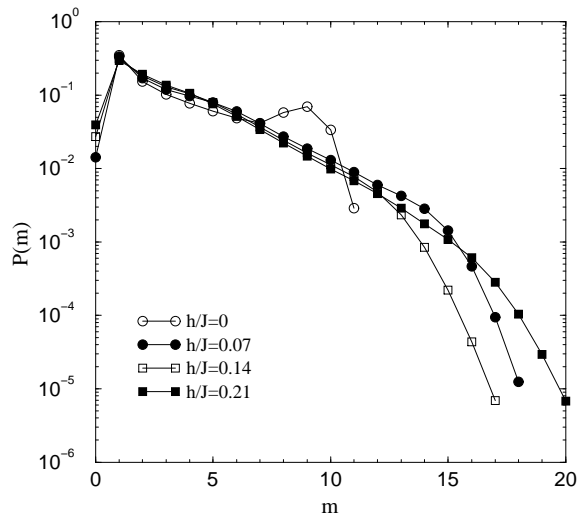}
\caption{Loops size distribution for $N=64$ chains at $\beta=128$ and 
different field strengths (Solution B simulations with $\epsilon=1/4$). 
$P(m)$ is the cumulative probability of loop sizes between $2^m$ 
($0$ for $m=0$) and $2^{m+1}-1$.}
\label{hist}
\end{figure}

Distributions of loop sizes at $\beta=128$ are shown in Fig.~\ref{hist} for
field strengths corresponding to magnetization plateaus ($h/J=0$ and $0.14$) 
and switching fields ($h/J=0.07$ and $0.21$). At $h=0$, there are no bounce 
processes and this appears to be reflected as a qualitatively different loop 
size distribution than for $h > 0$, with no very large loops and a larger 
probability of sizes in the range $2^8 - 2^{11}$. For all fields, there
is a quite sharp cross-over beyond which the probability becomes very small. 
Problems with loops that do not close \cite{pinaki,dorneich2} are therefore 
absent in this case. We did not have to impose any maximum size during the
loop construction in any of the simulations discussed in this paper.

In the studies of the 1D Heisenberg model in a field that we have
presented here, the new Solution B is clearly better than Solution A, 
although the difference is very large only for $h$ close to $0$ (but 
significant also for $h \to h_{\rm sat}$). Already with solution A 
the autocorrelation time for the magnetization is very short compared 
to other approaches. With the continuous-time worm algorithm 
$\tau_{\rm int}[M]$ is close to 100 even for system sizes as small as 
$N=10$ and $N=20$ \cite{kashurnikov}. 

\subsection{SSE simulations in 2D}

For the 2D XXZ-model (on periodic $L\times L$ lattices), we have calculated 
autocorrelation times versus the field strength in systems with isotropic 
couplings ($\Delta = 1$, $0 \le h \le h_{\rm sat} = 4J$), Ising-anisotropic 
systems in zero field ($\Delta \ge 1$, $h=0$), and the XY-model in zero 
and finite field ($\Delta =0$, $h/J=0,1/2$).

\begin{figure}
\includegraphics[clip,width=7.5cm]{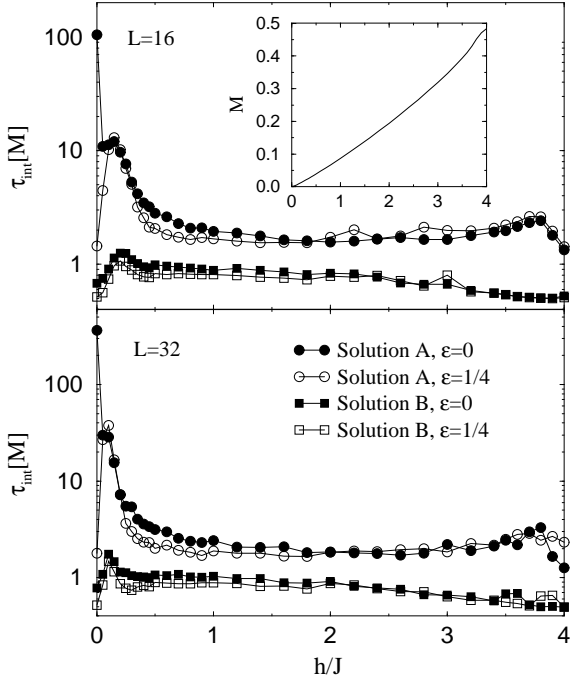}
\caption{Integrated autocorrelation times for the magnetization in
simulations of the 2D Heisenberg model in a magnetic field, using
Solution A (circles) and B (squares) and constants $\epsilon=0$ (filled
symbols) and $1/4$ (open symbols). The inset shows the magnetization
(the differences in $M$ between $L=16$ and $L=32$ are very small at the
inverse temperature $\beta=8$ used here).}
\label{htau2d}
\end{figure}

Fig.~\ref{htau2d} shows the field dependence of the autocorrelation time for 
the magnetization of $L=16$ and $L=32$ systems at inverse temperature 
$\beta=8$. With Solution A at $\epsilon=0$, a sharp drop in the 
autocorrelation time can be noted immediately when $h$ becomes non-zero. 
It is not surprising that the algorithm at $h=0$ is inefficient, since the 
only processes occurring here are the switch-and-reverse and the bounce 
(see Fig.~\ref{vpaths}). The bounce probability is high if it is not 
excluded ``by hand'', which would yield the much more efficient deterministic 
loop rules. With the bounce included, the actual closed loop is still 
deterministic but during its construction the propagating open end oscillates 
randomly back and forth along the {\it defacto} deterministic trajectory until
the loop closes or is annihilated via back-tracking all the way to the 
starting point. Once $h$ is non-zero, the loops become manifestly 
non-deterministic (since an additional vertex path becomes allowed)
and apparently, as seen in Fig.~\ref{htau2d}, even for a very small $h$ the 
simulation is much more efficient. This is in contrast to the 1D case (see 
Fig.~\ref{mltau}), where Solution A with $\epsilon=0$ is reasonably efficient 
even for $h=0$ and $\tau_{\rm int}[M]$ increases when $h$ is turned on. 
This difference between the 1D and 2D simulations may be related to the 
loop sizes (although the full explanation probably is more complex and
related to the different physical properties of the systems, which are
reflected in the loop structures). In 1D, the loops are relatively small, 
and for a small $h$ a large fraction of the constructed loops are then 
identical to the deterministic ones at $h=0$. In 2D the loops are much 
larger, and then even a small $h$ can allow most paths to ``escape'' from 
the $h=0$ deterministic loop trajectories so that there are not as many 
propagations back and forth along the same path as at $h=0$. Using a non-zero 
$\epsilon$ also makes the path non-deterministic, and Fig.~\ref{htau2d} shows 
very favorable effect of using $\epsilon=1/4$ in Solution A at $h=0$. For 
higher fields, there are only very minor advantages of a non-zero $\epsilon$, 
which is also in contrast to the 1D case. As in the 1D case, Solution B 
reduces the autocorrelation times very significantly at weak fields, and 
substantially also at higher fields. The differences between $\epsilon=0$ 
and $1/4$ in Solution B are small at all fields, however.

\begin{figure}
\includegraphics[clip,width=7.5cm]{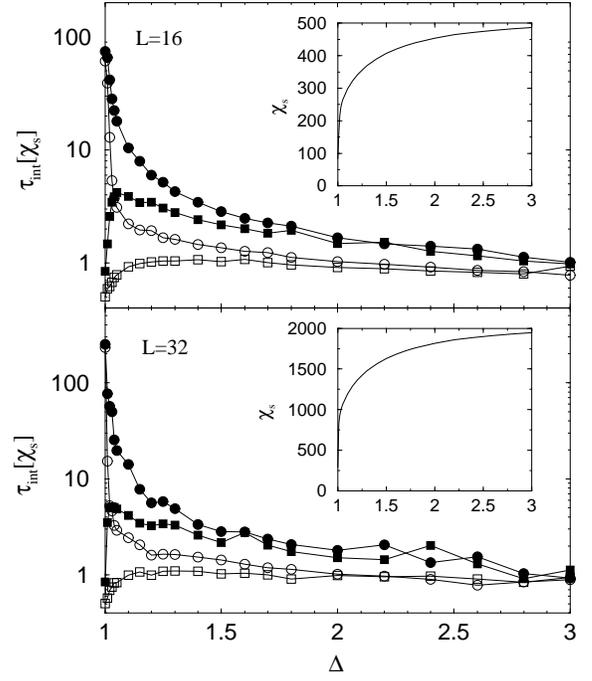}
\caption{Integrated autocorrelation times for the staggered susceptibility 
in simulations of the 2D anisotropic Heisenberg model at $\beta=8$. The 
symbols indicate Solutions A, B, and $\epsilon=0,1/4$ in the same way as in 
Fig.~\ref{htau2d}.}
\label{dtau2d}
\end{figure}

\begin{figure}
\includegraphics[clip,width=7.5cm]{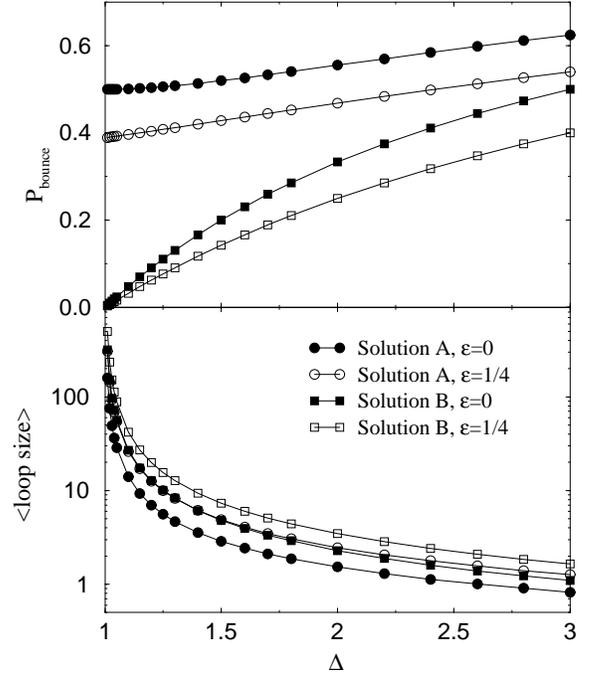}
\caption{Upper panel: Bounce probabilities in simulations of a $32\times 32$
anisotropic Heisenberg model at $\beta=8$. Lower panel: The average loop size
in the same simulations.}
\label{bounce2d}
\end{figure}

Fig.~\ref{dtau2d} shows autocorrelation times for the staggered susceptibility
of Ising-anisotropic systems in zero field at $\beta=8$. Solution B performs
significantly better than Solution A for $\Delta \alt 1.5$, but only marginally
better at higher $\Delta$. In this system $\epsilon = 0$ implies that the
closed loops are {\it defacto} deterministic for all anisotropies (the only 
allowed $\epsilon=0$ vertex processes are again the bounce and the 
switch-and-reverse). 
However, the symmetry of flipping and flipping back loops is 
broken when $\Delta > 1$ and the {\it defacto} fixed structure of the closed 
loops is not taken into account during their construction, neither with 
Solution A nor B (doing this would correspond to neglecting the bounces,
constructing a deterministic loop and then taking $\Delta$ into account in 
a Metropolis acceptance probability for actually flipping the loop, in a way 
analogous to what has been done with the standard world-line loop method for 
weak magnetic fields \cite{troyer}). Solution B minimizes the bounce 
probability and hence leads to more directed paths and, therefore, closing of 
the loops in fewer steps (and hence a larger number of completed loop in an 
MCS as defined here). Bounce probabilities are shown in the upper panel of 
Fig.~\ref{bounce2d}. When $\epsilon> 0$ the loops become manifestly 
non-deterministic, leading to 
significantly reduced autocorrelation times. The bounce probabilities are also
reduced, but for both solutions $P_{\rm bounce}$ still becomes large as 
$\Delta$ is increased. Nevertheless, the autocorrelation times continue to 
decrease. We do not expect this to be the case as $\Delta\to \infty$, where 
the model at fixed $\beta$ reduces to the classical Ising antiferromagnet 
at temperature $T\to 0$. In that limit, a classical single-spin flip would 
correspond to flipping spins on all SSE vertices on a given site (the number 
of which scales as $\beta\Delta$), which would be a slow process since the 
bounce probability is high. The lower panel of Fig.~\ref{bounce2d} shows 
that the average loop size becomes very small for large $\Delta$. The 
algorithm clearly does not reduce to a classical Swendsen-Wang or Wolff 
cluster algorithm as $\Delta\to \infty$ (in the classical algorithms the 
cluster size $\to N$ as $T\to 0$). However, at higher temperatures the 
algorithm could easily be supplemented with a cluster update which corresponds
exactly to the classical one (a multi-spin generalization of the flips of 
``free'' spins, where clusters of spins connected to each other by operators 
in $S_M$ can be flipped simultaneously without changing the weight if 
$\epsilon=0$ and $h=0$). As in the standard world-line loop algorithm
\cite{evertzchapter}, it is also possible to include loop-freezing in the 
deterministic operator-loop algorithm.

Note that there is essentially no structure in the Solution B autocorrelation 
time for $\epsilon=1/4$ in Fig.~\ref{dtau2d}, in spite of the fact that the 
scan over anisotropies should cross an Ising-type transition to an ordered 
state. At $\Delta=3$ the antiferromagnetic order is already at 
$\approx 97\%$ of the maximum (classical $T=0$) value, as can be inferred 
from the insets of Fig.~\ref{dtau2d} by using Eq.~(\ref{chisdef}). 

\begin{figure}
\includegraphics[clip,width=7.5cm]{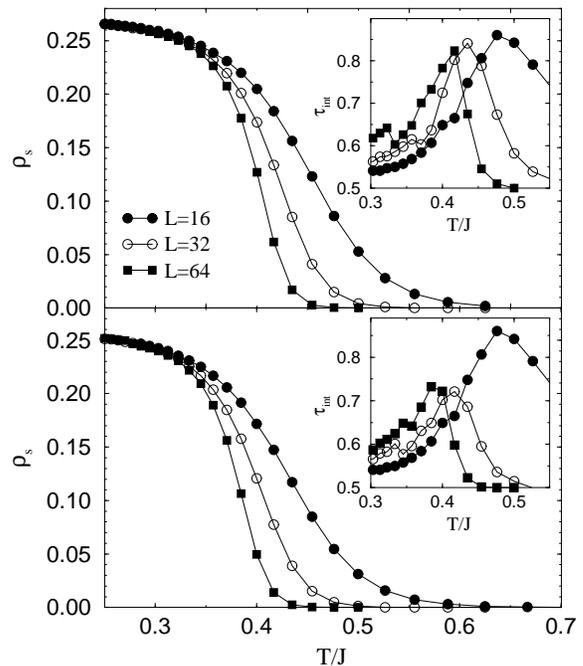}
\caption{Spin stiffness of the $XY$-model at zero external field (upper panel)
and at $h/J=0.5$ (lower panel). The insets show the corresponding integrated 
autocorrelation times. Solution B with $\epsilon=\epsilon_{\rm min}$ (see
Table I) was used in all cases.}
\label{xyrho}
\end{figure}

For the XY-model ($\Delta =0$), the directed loop equations
have a solution without bounces for all fields up to the saturation field.
We find that the resulting algorithm is very efficient, with autocorrelation
times smaller than one for all system sizes and temperatures that we have 
studied. Fig.~\ref{xyrho} shows results for the spin stiffness as
a function of temperature for zero field as well as at $h/J = 0.5$. The
corresponding autocorrelation times are peaked around the Kosterlitz-Thouless
(KT) transition temperature but do not grow with the system size. The
KT transition in the $h=0$ system has been studied to high accuracy using a 
continuous-time world-line loop algorithm, with the result $T_{\rm KT}/J 
\approx 0.342$ \cite{harada}. Our $h=0$ data are in complete agreement with 
the previous results. We find that the data for $h=0.5$ shown in 
Fig.~\ref{xyrho} can be collapsed onto the $h=0$ data if $T$ and 
$\rho_s$ are both scaled by the same factor ($\approx 1.05$ for $h/J=0.5$), 
in accord with the universality of the transition. More extensive results 
for this model will be presented elsewhere.

\subsection{PIM simulations}

Next we will show some results for autocorrelation times obtained using the 
PIM implementation of the directed loop algorithm. To make a reasonable 
comparison with the autocorrelation times for the SSE, we will also in the 
PIM define a MCS so that it includes $N_l$ loops, where $N_l$ is determined 
such that on average the total path length, excluding the first path segment 
immediately following each bounce, of all $N_l$ loops in an MCS is equal to 
$\beta N$; the space-time volume (in the PIM, each path segment has a length
in imaginary time, in contrast to the SSE where the steps are just 
counted). This definition is chosen so that it corresponds reasonably closely 
to the definition used in the SSE. However, it could be argued that a better 
definition of the total path length would be to add all the path segments
but instead of excluding the segment immediately following a bounce one 
would subtract the part of the path immediately following a bounce that
overlaps with the path segment preceding the bounce (with special care 
taken for consecutive bounces). This would more 
accurately take into account the fraction of spins actually flipped. We have 
here used the first definition of the MCS as it corresponds more closely 
to how we define an MCS in the SSE method (where a different treatment of the 
bounces could of course also be implemented---see Sec.~II D). 

\begin{figure}
\includegraphics[clip,width=7.5cm]{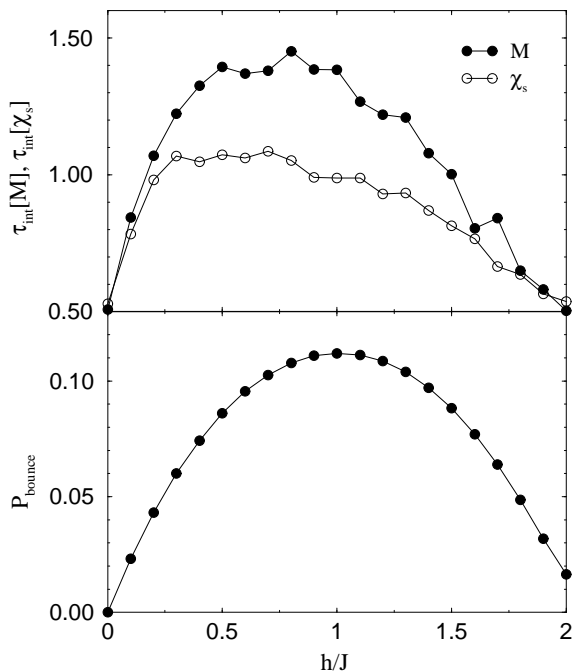}
\caption{Upper panel: Integrated autocorrelation times vs external field for 
the magnetization (solid circles) and staggered susceptibility (open circles) 
in PIM simulations of an $N=64$ Heisenberg chain at $\beta=16$. 
Lower panel: Bounce probability vs external field in the same simulation.} 
\label{pim_n64}
\end{figure}
 
Generally speaking the computer implementation of the PIM is more complex 
than SSE, as it is always necessary to keep track of the spin states on 
neighboring sites in the PIM. This is not required in the SSE formulation,
where the vertices contain all the information needed. Therefore our computer 
code for the PIM is not as efficient as the SSE code in generating a single 
MCS, and so we will be content in this section to show just a few PIM 
autocorrelation results. As Solution A of the directed loop equations 
was already shown above to be much less effective than Solution B, we will 
in this section just show results for Solution B. 

Fig.~\ref{pim_n64} shows the integrated autocorrelation times 
$\tau_{\rm int}[M]$ and $\tau_{\rm int}[\chi_s]$ for a 64-site Heisenberg 
chain ($\Delta=1$) at inverse temperature $\beta=16$ as functions of 
the magnetic field. Comparing with the SSE results in Figs.~\ref{l64_m} 
and \ref{l64_x}, it is seen that $\tau_{\rm int}[M]$ is comparable to the 
$\epsilon=1$ case while $\tau_{\rm int}[\chi_s]$ is more similar to the 
$\epsilon=0$ curve, except close to $h=0$ where it also behaves more like 
the $\epsilon=1$ case. The lower panel of Fig.~\ref{pim_n64} shows the field 
dependence of the bounce probability. $P_{\rm bounce}$ is here defined as 
the number of bounces divided by the total number of times the path building 
changes, either by moving to a neighbor site or by back-tracking. This measure
is not directly comparable to the definition in the SSE case, as the moves 
$c$ and $c^\prime$, where the path continues on the same site, are not counted
in the denominator of $P_{\rm bounce}$ (they are infinitely many in the PIM). 
Nevertheless, the general behavior of $P_{\rm bounce}$ versus $h$ is the 
same for the two methods.   

\begin{figure}
\includegraphics[clip,width=7.5cm]{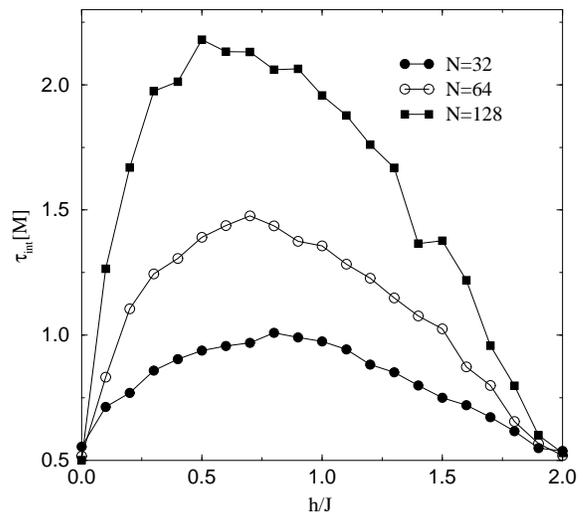}
\caption{Field dependence of the integrated autocorrelation times for the 
magnetization in PIM simulations of chains of different lengths $N$ at inverse 
temperature $\beta=N/4$.}
\label{pim_magall}
\end{figure}

In Fig.~\ref{pim_magall} we have plotted $\tau_{\rm int}[M]$ as a function of 
magnetic field for different chain sizes N. In all cases $\beta = N/4$. As 
in the SSE case (Fig.~\ref{mltau}) we see an increase in $\tau_{\rm int}[M]$ 
with system size for small to intermediate fields. However, the maximum PIM 
autocorrelation times are about $50\%$ smaller than in the SSE 
$\epsilon=1/4$ case. 

\begin{figure}
\includegraphics[clip,width=7.5cm]{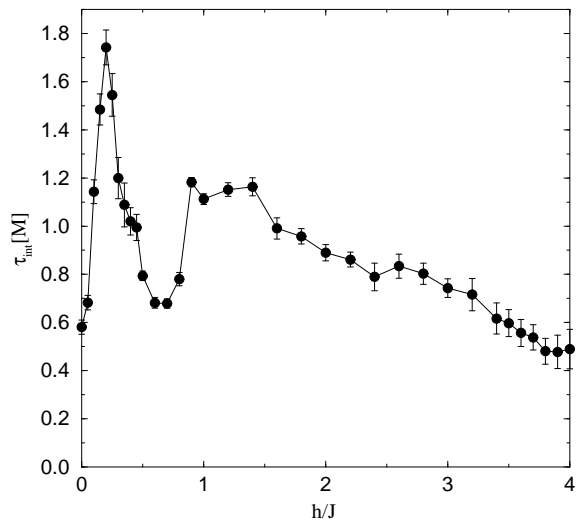}
\caption{Integrated autocorrelation times (PIM) vs external field for the 
magnetization of an $16 \times 16$ Heisenberg square lattice at $\beta=8$.} 
\label{pim_mag2d}
\end{figure}

We have also carried out simulations of the 2D Heisenberg model using the 
PIM. In Fig.~\ref{pim_mag2d} we show results for $\tau_{\rm int}[M]$ for a 
$16\times 16$ lattice at $\beta=8$. Here the behavior is almost identical 
to the SSE results shown in Fig.~\ref{htau2d}, where there is only a small
dependence on $\epsilon$.

From these examples it can be seen that the PIM generally has shorter 
autocorrelation times than SSE in cases where the SSE results show a
significant dependence on the constant $\epsilon$. In some sense the PIM 
corresponds to the $\epsilon \to \infty$ limit of SSE, as in this limit 
the continue-straight processes also dominate the loop construction
in SSE. In cases where the SSE autocorrelations converge slowly to their 
$\epsilon=\infty$ limit the PIM approach may hence be more efficient (since
in SSE the computation time for one MCS grows linearly with $\epsilon$ in
this limit). However, in assessing a method's efficiency one should also 
take into account the cost of performing a single MCS. This of course depends 
heavily on the actual computer implementation of the directed loop 
algorithm. That is, what kind of data structures are used to represent the 
spin and vertex configurations, what kind of search algorithms are used for 
finding spin states at a given time in the PIM, e.t.c. While we do not attempt
to compare the PIM and SSE in this respect here, it is quite clear that it 
is often easier to find a fast and effective implementation for the SSE than 
for the PIM. We also note that the convergence to the $\epsilon\to \infty$
limit in the SSE is relatively fast in all cases we have studied so far. 
The convergence appears to be slowest in 1D, but even there the reduction 
of the autocorrelation times becomes small beyond $\epsilon=1$, where
they are similar to the PIM autocorrelations.

\subsection{Dynamic exponent}

An interesting question is how the autocorrelation time diverges with the 
system size in simulations at a critical point. The 1D Heisenberg model at
$h=0, T=0$ exhibits power-law ($1/r$) decay of the staggered spin-spin 
correlation function and is a hence a quantum critical system \cite{luther}. 
We have studied the integrated autocorrelation time for the staggered spin 
susceptibility in this model as the system size $N$ is increased and the 
inverse temperature $\beta=N/4$. The staggered susceptibility should couple 
to the slowest mode of the simulation, and its autocorrelation time is 
therefore expected to diverge asymptotically according to a power law;
\begin{equation}
\tau_{\rm int}[\chi_s] \sim \beta^{z},
\end{equation}
where $z$  is the dynamic exponent of the simulation. Note that it is here 
essential that $\beta$ and $N$ are taken to infinity at a fixed ratio (as the 
physical dynamic critical exponent relating space and imaginary time is $1$). 
It is interesting to compare SSE simulations with Solution B at different 
$\epsilon$-values (we do not consider solution $A$ here since it is much
less efficient than Solution B). It is also interesting to compare 
the two possible ways of flipping the loops when $\epsilon=0$. At $h=0, 
\epsilon=0$, Solution $B$ reduces to the deterministic 
operator-loop \cite{sse3}. As discussed in Sec.~II D, instead of
constructing a fixed number of loops per MCS at random, all loops can then be 
constructed and flipped independently of each other with probability $1/2$. 
This is analogous to the Swendsen-Wang \cite{swendsen} algorithm for the 
classical Ising model. For the Ising model, it is known that it is more 
efficient (i.e., $z$ is smaller) to construct the clusters one-by-one using 
the Wolff algorithm \cite{wolff}. 

\begin{figure}
\includegraphics[clip,width=8cm]{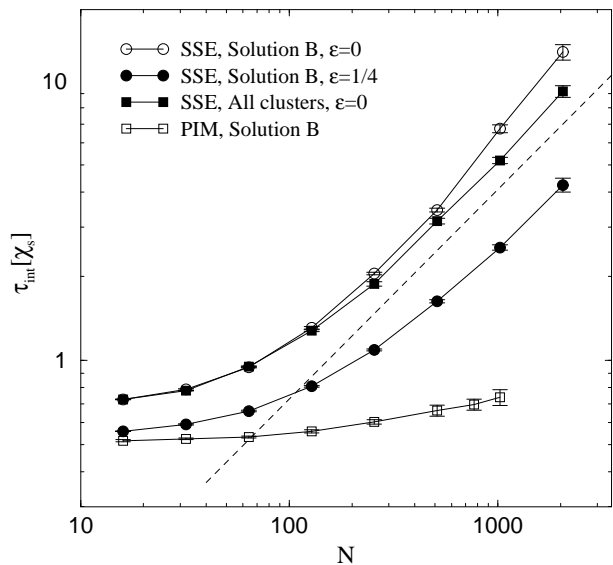}
\caption{Autocorrelation times for the staggered susceptibility of the 
isotropic Heisenberg chain at $\beta=N/4$ obtained in SSE and PIM simulations 
with Solution B. The dashed line has slope $0.75$.}
\label{logtau1d}
\end{figure}

In Fig.~\ref{logtau1d} we show results of Solution B simulations with 
$\epsilon=0$ and $1/4$ along with results from $\epsilon=0$ simulations were 
all clusters were constructed. The autocorrelation times of the two 
$\epsilon=0$ simulations are very similar, but for large systems marginally 
shorter when all clusters are constructed. Hence, there is here no advantage 
in constructing the clusters one-by-one. This is most likely related to the 
fact that in order to change the loop structure in the SSE simulations at 
$h=0,\epsilon=0$, diagonal updates also have to be carried out. In the scheme 
used here, diagonal updates are only performed at the beginning of each MCS, 
and hence the same loop can be constructed several times in one MCS if they
are constructed at random. It is then more efficient to construct all loops 
once. In order to achieve an advantage similar to the Wolff algorithm, one 
would have to construct a new scheme for the diagonal updates, which certainly
could be possible but which we have not yet attempted. As in the other cases 
we have discussed above, there is a significant improvement when 
$\epsilon=1/4$ is used in Solution B. However, the dynamic exponent appears to
be the same in all cases; $z \approx 0.75$. In Fig.~\ref{logtau1d} we also 
show PIM results. It is clear that the autocorrelation times here are 
significantly shorter but most likely the dynamic exponent is the same as 
in SSE. The shorter PIM autocorrelation times are consistent with the 1D
results shown above in Secs.~V A and C, and clearly we could also reduce 
the SSE autocorrelations by increasing $\epsilon$ further.

\begin{figure}
\includegraphics[clip,width=8cm]{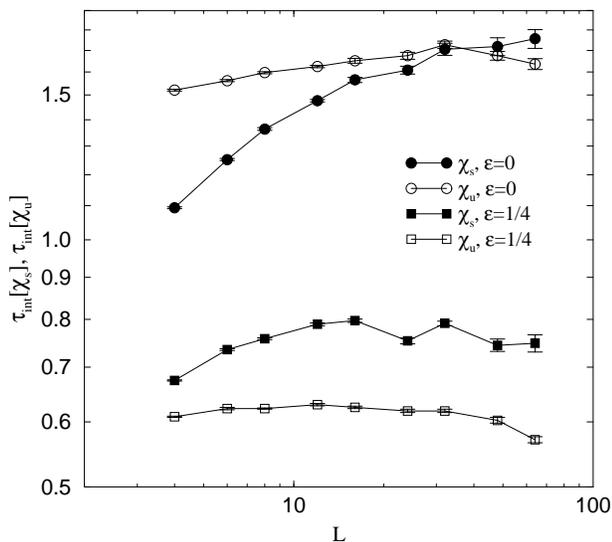}
\caption{SSE autocorrelation times for the staggered and uniform susceptibility
of the Heisenberg bilayer close to its quantum critical point 
($J_\perp/J=2.5225$ was used). The inverse temperature $\beta=L/J$.}
\label{logtau2d}
\end{figure}

In 2D, a well studied quantum critical system is the Heisenberg model on
two coupled layers (bilayer), with intra-plane coupling $J$ and inter-plane 
coupling $J_\perp$ \cite{hida}. The $T=0$ antiferromagnetic long-range order 
in this model vanishes at a critical inter-plane coupling $(J_\perp/J)_c 
\approx 2.525$ \cite{ssespin2}. Some autocorrelation results for both SSE 
and PIM simulations of this model at $J_\perp/J = 2.524$ have been presented 
recently \cite{dorneich2} and indicate that the dynamic exponent $z \approx 0$
in both methods. Our most recent simulations indicate that $(J_\perp/J)_c 
\approx 2.5225$, i.e., slightly lover than the previous estimate 
\cite{ssespin2}. In Fig.~\ref{logtau2d} we show integrated autocorrelation 
times for several quantities at this coupling, using both $\epsilon=0$ 
and $1/4$ in Solution B simulations. In the $\epsilon=0$ case, all clusters 
were constructed in each MCS. We again note significant shorter 
autocorrelation times in the non-deterministic simulation ($\epsilon =1/4$). 
However, the deterministic simulation is significantly faster. One 
deterministic MCS at $\epsilon =0$ typically only requires $50-60\%$ of the 
CPU time of a generic Solution B MCS at $\epsilon =1/4$. The net gain in 
simulation efficiency with $\epsilon > 0$ is therefore only marginal 
in this case. All our results are consistent with $z=0$, although with 
$\epsilon=0$ the convergence to a size-independent behavior is rather 
slow. We have not carried out PIM simulations of this system.

\begin{figure}
\includegraphics[clip,width=8cm]{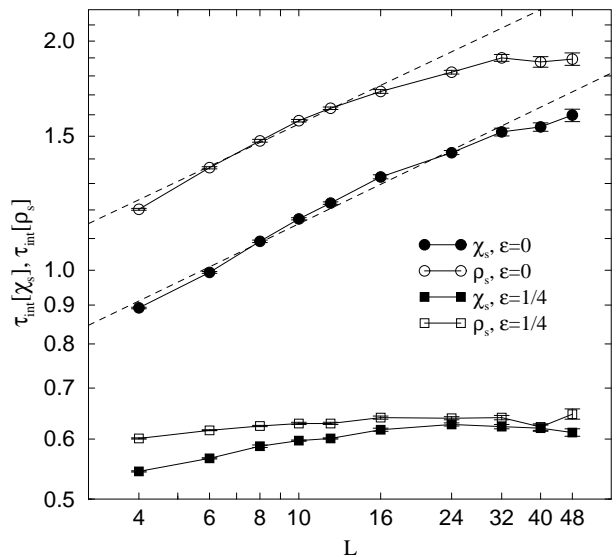}
\caption{Autocorrelation times for the staggered susceptibility and the
spin stiffness of the 3D Heisenberg model close to its critical temperature 
($T/J=0.944$ was used). The lines correspond to scaling $\sim L^{1/4}$.} 
\label{logtau3d}
\end{figure}

Next we consider the 3D Heisenberg model, which undergoes a phase transition
to an antiferromagnetic state at a non-zero temperature \cite{rushbrooke}.
According to recent SSE simulations, using systems with $N=L^3$ sites and
$L$ up to $16$, the critical temperature $T_c/J = 0.946 \pm 0.001$
\cite{aws3d}. These simulations were carried out using only local updates. 
With the operator-loop update, much larger systems can be studied. We have 
carried out simulations for $L$ up to $48$ close to the critical temperature. 
Based on the results, we believe that $T_c$ is at the low end of the previous 
estimate, likely very close to $0.944$. Fig.~\ref{logtau3d} shows 
autocorrelation times for the staggered susceptibility and the spin 
stiffness at $T/J=0.944$, obtained using the deterministic SSE algorithm 
with $\epsilon=0$ (constructing all clusters during each MCS) and Solution 
B with $\epsilon=1/4$. Here the $\epsilon=0$ results are initially consistent
with a dynamic exponent $z \approx 0.25$, but for the largest sizes there
seems to be a change in behavior, possibly a convergence corresponding to
$z = 0$. The $\epsilon=1/4$ simulation is fully consistent with $z=0$.

\section{Low-field Magnetization of the 2D Heisenberg model}

As an example of an application made possible with Solution B of the 
directed loop equations, we here present SSE simulation results for the 2D 
Heisenberg model in a weak magnetic field. At very low temperatures, the 
field dependence of the magnetization exhibits a step-structure due to the 
gaps between the lowest-energy states with magnetization 
$m_z=0,\pm 1,\pm 2,\ldots \pm N/2$. These gaps can be extracted from the 
calculated magnetization curve. For the isotropic Heisenberg model, the gaps 
are exactly the gaps between the degenerate spin multiplets with total spin 
$S=0,1,\ldots$ in the absence of the field. 

In an antiferromagnetically ordered system, such as the 2D Heisenberg model, 
the energies of the $S > 0$ multiplets relative to the $S=0$ ground state 
should correspond to the excitations of a quantum rotor when $S \ll \sqrt{N}$.
The over-all energy scale can be related to the uniform (transverse) magnetic 
susceptibility \cite{rotor}:
\begin{equation}
E(S) = {S(S+1)\over 2L^2 \chi_\perp},
\label{rspectrum}
\end{equation}
where $L^2 = N$. The asymptotic validity of this 
relation has been verified using quantum Monte Carlo estimates for small 
$S$ and $L$ up to $16$  \cite{runge,lavalle}. Recently, a slow convergence 
of the spectrum for $S \sim L$ was been pointed out for the 2D Heisenberg 
model with spin-$1/2$ \cite{lavalle}. A systematic study of $E_L(S)$ for 
systems larger than $L=16$ was not possible, however, because of the large 
statistical errors in the energy differences.

With the directed loop algorithm we can instead extract the energy gaps
using the field dependence of the magnetization. As we have shown in
Sec.~V, the new Solution B shortens the autocorrelation times very
significantly for low fields, which is what we need in order to accurately
extract the energy levels for $S$ ranging from $0$ to $\sim L$. We will 
present our complete results of such calculations elsewhere. 
Here we will demonstrate the power of the new method by focusing on the first 
few levels for system sizes $L$ up to $64$, i.e., the number of spins is 
$16$ times larger than in the previous studies \cite{runge,lavalle}.  

\begin{figure}
\includegraphics[clip,width=8cm]{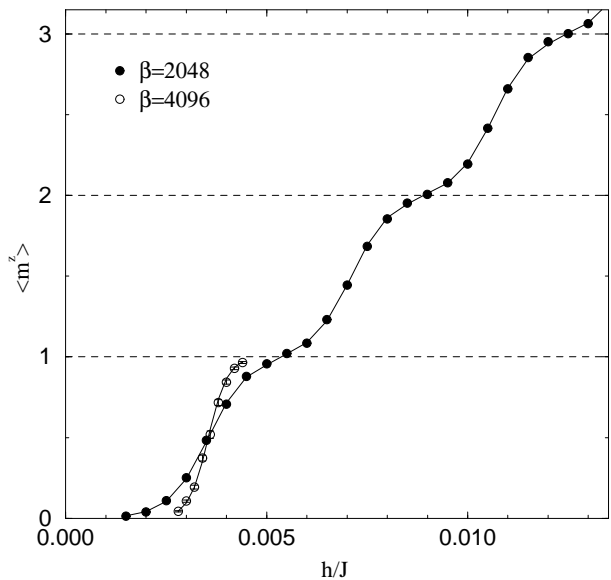}
\caption{Total magnetization vs external field in the 2D Heisenberg
model with $L=64$ at two inverse temperatures. The curves were calculated
using four fitted energies $E_L(S)$ (the same for both curves).} 
\label{m64}
\end{figure}

\begin{figure}
\includegraphics[clip,width=8cm]{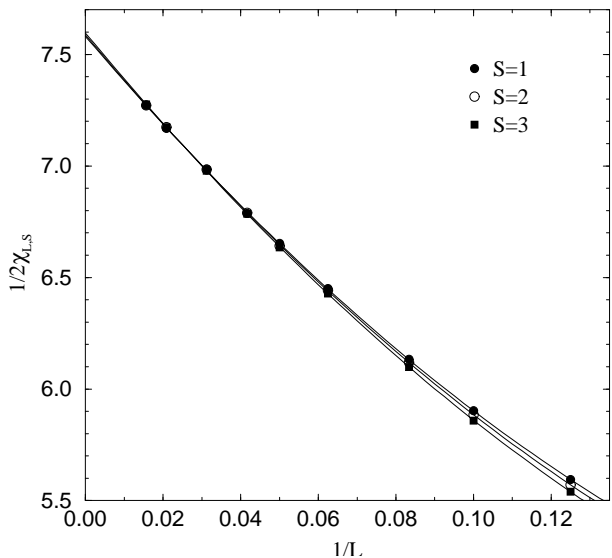}
\caption{Inverse susceptibility extracted using the energies of the
$S=1,2$, and $3$ multiplets. The curves are quadratic fits.}
\label{xl}
\end{figure}

In order to see the step-structure needed to extract the energy levels 
$E_L(S)$ for small $S$, the temperature has to be below the $S=1$ gap,
which according to Eq.~(\ref{rspectrum}) and previous estimates of the 
susceptibility ($\chi_\perp \approx 0.065/J)$ is approximately $0.004/J$
for $L=64$. In practice, we have used inverse temperatures $\beta$
corresponding to roughly $1/10$ of the gap. We have fitted the numerical 
results to a magnetization curve $\langle m_z\rangle$ calculated using 
energy levels of the form
\begin{equation}
E_L(S,m_z) = E_L(S) - hm_z, ~~~ m_z = 0,\pm 1,\ldots \pm S,
\end{equation} 
at the same temperature as in the simulations. We adjust the energies
$E_L(S)$ to give the best match between the calculated and theoretical
magnetization curves. Fig.~\ref{m64} shows results for $L=64$ at $\beta=2048$ 
and $4096$. We used the same fitted levels $E_L(S)$ at both 
temperatures (clearly, the $S=1$ level completely dominates the $\beta=4096$ 
results, which include only the first magnetization step). As in
Ref.~\onlinecite{lavalle}, we define a spin- and size-dependent 
susceptibility using the energy levels $E_L(S)$ obtained in this
fitting procedure;
\begin{equation}
{1\over 2\chi_{L,S}} = {L^2 E_L(S) \over S(S+1)},
\end{equation}
and extrapolate data for fixed $S$ to infinite size in order to determine the 
thermodynamic susceptibility $\chi_\perp$. Fig.~\ref{xl} shows 
our results for $S=1,2,3$ and system sizes ranging from $L=8$ to $L=64$. The
results up to $L=16$ agree very well with those presented previously 
\cite{lavalle}, but our statistical errors are considerably smaller. The 
collapse of the three curves onto each other for large systems demonstrate 
the validity of Eq.~(\ref{rspectrum}) for small $S$. Extrapolating the three 
data sets to infinite size gives the susceptibility $\chi_\perp = 0.0659 \pm 
0.0002$, again in good agreement with Ref.~\onlinecite{lavalle} but with 
a considerably reduced statistical error.

For the $L=64$ simulations at $\beta=4096$, the CPU time needed to perform
one MCS is approximately 40s on an Intel Pentium III running at 866 Mhz.
The results shown in Fig.~\ref{m64} are based on $3-8 \times 10^4$ MCS
for each data point.

\section{Summary and Discussion}

We have introduced the concept of directed loops in stochastic series 
expansion and path integral quantum Monte Carlo and implemented them for 
simulations of the $S=1/2$ XXZ-model in an external magnetic field. The 
directionality of the loop reflects the asymmetry between the operation of 
flipping the spins along the loop and the reverse operation of flipping 
back those spins. Such an asymmetry is not present in the standard world-line
loop algorithm \cite{beard,evertz,evertzchapter}, which as a consequence is 
restricted to certain regions of parameter space. 
Quite generally, there is a hierarchy of three classes of directed loops.
In the most general case the loop can back track during construction. In
some regions of the parameter space the back tracking can be excluded,
and in some further restricted regions the loops become symmetric
(non-directional) and reduce to the type of loops previously considered
for world-line \cite{beard,evertz,kawashima} and SSE \cite{sse3} simulations. 
Hence, the directed
loop framework constitutes a natural generalization of the loop-cluster
concept \cite{evertz}. We have shown that the transitions between the different
levels of the hierarchy can be made smooth by minimizing the probability
of back-tracking when solving the directed loop equations.
We have also demonstrated that the algorithm based on this solution works 
very well in the full parameter space of the XXZ-model. 

Our scheme appears to be much more
efficient than the worm algorithm for continuous-time path integral 
simulations \cite{prokofev}, which also is applicable in the full parameter
space but does not exhibit the three-level hierarchy of the directed loops
(at least not in its current formulation). The
configuration space involving two moving discontinuities was used first in
the worm algorithm for the purpose of measuring off-diagonal correlation
functions (Greeen's functions). It was also the first method that was
practically useful in the presence of external fields. It is, however, not
the presence of the discontinuities that makes the worm algorithm and SSE
operator-loop \cite{sse3} algorithm applicable in the presence of external
fields. One can also think of the construction of the standard world-line
loops \cite{evertz,evertzchapter} in terms of moving discontinuities, but they are more
constrained in their motion and therefore cannot take external fields
into account. Hence, it is the rules for moving the discontinuities
that determine whether or not a loop or worm simulation is efficient.
The directed loop equations constitute a framework for optimizing
these rules. Below we will comment on the similarities and differences between worms and directed loops.

The operator-loop update previously constructed for SSE simulations \cite{sse3}
corresponds to a particular solution (A) of the directed loop equations.
We have here constructed a different solution 
(B), which minimizes the probability of back tracking in the loop construction
and therefore is more efficient. The new solution B completely eliminates 
back-tracking (bounce processes) in the XXZ-model for $z$-anisotropies 
$-1 < \Delta < 1$ up to a finite external field $h$ (up to the saturation 
field for $\Delta=0$ and only exactly at $h=0$ for $|\Delta| =1$). 
In other interesting parameter regions the bounce probability is typically
a few percent or less. Our simulation results show that the new solution can 
decrease the autocorrelation times by up to an order of magnitude or more 
in cases where Solution A is the least efficient (at weak and
intermediate magnetic fields and anisotropies). The algorithmic discontinuity 
of the previous approach (which amounted to using a very efficient 
deterministic algorithm at $h=0$ and the much less efficient generic 
Solution A for $h > 0$) is hence avoided with Solution $B$, where the bounce 
probabilities and the autocorrelation times smoothly connect to those of the 
deterministic algorithm. However, our results also indicate that the 
deterministic loop construction at $h=0$ is not always the most efficient. 
With a non-deterministic solution (Solution $B$ with the constant $\epsilon 
> 0$ in the bond operator) the operator paths becomes more random, which has a 
favorable effect on the autocorrelations.

In addition to being more efficient in terms of the autocorrelation times
measured in units of our defined MCS, Solution B is also typically faster
as the number of operations required to perform one MCS is smaller (because
of the smaller bounce probability). In terms of ease of implementation,
Solution A is more straight-forward as it is directly given in terms of 
matrix elements of bond-operators. In order to implement Solution B for a 
new Hamiltonian, one first has to investigate the subclasses of vertices 
with their directed loop segments and then minimize the bounce probabilities 
for all non-equivalent classes. SSE with Solution A (and other special
solutions for Heisenberg and XY-models) have already been used for a number 
of different lattices and Hamiltonians \cite{ssexy,ssespin1,ssespin2,wessel1,perc,wessel2,yunoki,hebert,dorneich1,schmid,torsten,pinaki,henelius2}, 
but so far we 
have only investigated Solution B for the XXZ-model discussed in this paper. 
We expect generalizations to a wide range of other models to be relatively
straight-forward.

In the continuous-time path integral, Solution B of the directed loop 
equations for zero field and $|\Delta| \le 1$ results in an algorithm 
identical to the standard world-line loop algorithm \cite{beard,evertzchapter}.
The generic algorithm, which includes a probability of back-tracking as the
loop is constructed, has some features in common with the worm algorithm
\cite{prokofev}. The extended configuration space with an open world-line 
segment (the worm) is the same in the two methods (and is analogous also in 
the SSE operator-loop construction, although the representation there is
discrete rather than continuous). However, there are important differences 
in the actual processes used to propagate the path (or worm). In the worm 
algorithm the ``jump'' and ``reconnection'' procedures involve the creation 
or annihilation of a kink, in which one of the worm ends jumps from one site 
to another and spins are flipped on finite equal-length segments of imaginary 
time at both the initial and final sites \cite{prokofev}. The location in 
time of the worm end does not change in these processes, but is accomplished 
in separate updates. In the PIM directed loop scheme, the movement in imaginary
time and the creation (or annihilation) of a kink is combined, and in each 
step spins are flipped on a finite segment of imaginary time at a single site 
only. This dynamics follows naturally from the vertex-representation 
introduced for the SSE operator-loop algorithm \cite{sse3}, where a single 
spin is flipped on a link connecting two vertices and the possible sites 
(the same or a specific neighbor site) and direction (forward or backward) 
for the next step is dictated by the four legs of the vertex. Here we have 
directly translated this dynamics into the path integral simulation by 
borrowing ides from the continuous-time loop algorithm \cite{beard}. The 
simulation dynamics is hence different from the worm algorithm, and the worm 
algorithm does not correspond to a solution of the directed loop equations. 
Our autocorrelation results show that the directed loop scheme is much more 
efficient than the worms in simulations of the Heisenberg chain in a magnetic 
field, for which our measured autocorrelation times for small systems are 
almost two orders of magnitude smaller than those reported  for the worm 
algorithm \cite{kashurnikov}. We expect the superior performance of the
directed loop scheme to be quite general, as the bounce minimization
achieved with Solution B has no counterpart in the worm algorithm (although
it may be possible to develop a generalization). There are, however, very 
interesting aspects of the worm scheme which could also perhaps be 
incorporated for the directed loops, e.g., the space-time potential 
introduced in order to more efficiently measure Green's functions at 
long distances \cite{prokofev}.

Comparing implementations of the directed loops within the SSE and PIM
representations, one difference is that in the former there is an adjustable 
parameter $\epsilon$ (a constant added to the bond Hamiltonian operators)
which is not present in the latter. We have noted that a non-zero 
$\epsilon$ has generally favorable effects on the autocorrelations in the 
SSE, but a large value is not practical since the computation time also 
increases with $\epsilon$. In some sense, the PIM corresponds to SSE with 
$\epsilon \to \infty$, and one might therefore expect the PIM implementation to
be more efficient. However, in practice the opposite is often true since 
already a small $\epsilon$ in the SSE can give autocorrelation times close to 
the $\epsilon \to \infty$ limit, and the computation time for one MCS 
can be significantly shorter in SSE. PIM algorithms should be more 
efficient in cases where the diagonal part of the Hamiltonian dominates in 
the internal energy, as the PIM configurations (which do not contain diagonal 
operators) then are smaller than the corresponding SSE configurations
\cite{irsse}. Another important aspect is the ease of implementation and 
optimization of the simulations for various models. We have found the discrete
nature of the SSE configuration space, where the vertices locally contain all
information needed to construct the loops, to be a distinct advantage in 
this respect.

An interesting question is whether the directed-loop approach could be used 
to further extend the applicability of the meron concept \cite{chandrasekharan}
for solving sign problems. We have shown that for the XXZ model back-tracking
in the loop construction can be avoided in a larger region of the parameter 
space than where the loop-algorithms previously used for studying merons are 
applicable (specifically, at non-zero external fields in XY-anisotropic 
systems). The possibility of generalizing the meron concept to the whole 
non-back-tracking region should be investigated.

\begin{acknowledgments}
We thank K. Harada and N. Kawashima for pointing out an error in region 
$\rm V$ of Table \ref{bounceweights} and Figure \ref{phasediagram} in an 
earlier version of this article. 
AWS would like to thank P. Henelius and M. Troyer for 
discussions. Both authors acknowledge support from a Nordic network project 
on Strongly Correlated Electrons at NORDITA. AWS also acknowledges support 
from the Academy of Finland (project 26175) and the V\"ais\"al\"a Foundation.
\end{acknowledgments}

\appendix

\section{Program Implementation of the SSE Method}

The computer implementation of a simulation method can of course be done
in several different ways and is an issue more technical in nature than
the mathematical definition of the underlying algorithm. Nevertheless, for 
the benefit of readers wishing to quickly construct a simple but efficient 
simulation program, we here briefly outline the basic aspects of our 
implementation of the SSE algorithm with the operator-loop update.
Some programs are also available online\cite{www}.

We first introduce the main data structures used to store the SSE 
configuration in computer memory. The state $|\alpha \rangle$ is stored as 
$Spin[s]=\pm 1$ representing the up and down spins at the sites $s$ 
($s=1,\ldots,N$). The operator-index sequence $S_M$ can be packed into an 
array $Sm[j]$ ($j=0,\ldots ,M-1$), with $Sm[j]=2b$ and $Sm[j]=2b+1$ 
($b=1,\ldots ,N_b$) corresponding to diagonal and off-diagonal bond-$b$ 
operators, respectively, and $Sm[j]=0$ representing fill-in unit operators. 
The lattice geometry can be coded into a list of sites $i(b),j(b)$ connected 
by the bonds $b$, i.e., $Site[1,b]=i(b)$, $Site[2,b]=j(b)$. The linked 
vertices are stored in the form of two lists, one containing the links and 
one the vertex types. The vertex types $Vtx[p]=1,\ldots,6$ ($p=0,\ldots, n-1$)
correspond to the six vertices shown in Fig.~\ref{vertices}. The links 
$Link[j]$ ($j=0,\ldots,4n-1$) are arranged such that $Link[4p+i]$ 
($p=0,\ldots,n-1$, $i=0,1,2,3$) contains the link (which is an integer
referring to another element in $Link$) for leg $i+1$ of vertex $p$ [the 
leg numbers $1,2,3,4$ are defined in Eq.~(\ref{flipvtx})]. The 
double-linked nature of the list implies that if $Link[a]=b$ 
then $Link[b]=a$. 

The diagonal update is straight-forward: For $j=0,\ldots,M-1$, a bond $b$
is generated at random for each $Sm[j]=0$, which is changed to $Sm[j]=2b$
with the probability (\ref{diap1}). If the change is made, the number of
bond-operators present increases by $1$, i.e., $n \to n+1$. For each diagonal
element, i.e., $Sm[j] > 0$ and even, the change to $Sm[j]=0$ and $n \to n-1$ 
is carried out with the probability (\ref{diap2}), where $b=Sm[j]/2$. If 
$Sm[j]$ is an odd integer it corresponds to an off-diagonal operator at 
bond $b=Sm[j]/2$ and the corresponding spin states should propagated, i.e.,
for $a=1,2$, $Spin[Site[a,b]] \to -Spin[Site[a,b]]$, in order for the matrix 
elements in Eqs.~(\ref{diap1}) and (\ref{diap2}) to be available as needed.

To understand the implementation of the linked vertex list, it is useful 
to keep in mind Fig.~\ref{linked} and the numbering of the vertex legs
exemplified in Eq.~(\ref{flipvtx}). In order to construct the lists $Link$ and
$Vtx$, two temporary arrays $First[s]$ and $Last[s]$ ($s=1,\ldots,N$) are 
needed. $First[s]$ will contain the first vertex leg on site $s$, i.e., 
$First[s]=4p+i$ means that the first operator acting on site $s$ is 
the $p$:th bond-operator $(p=0,\ldots ,n-1)$ in $Sm$ and the vertex leg acting
on the site is $l=i+1$ (where $l$ will always be  $1$ or $2$, as these are
the legs before the operator has acted). In an analogous way, $Last[s]=
4p+i$ refers to the last operator acting on site $s$ (where $l=i+1$
now will always be $3$ or $4$, since these are the legs after the operator 
has acted). All elements are initialized to $First[s]=Last[s]=-1$ before the 
construction of the linked list starts. Whereas $First[s]$ will be set at 
most once (never if no operator acts on site $s$), $Last[s]$ can be updated 
several times as the operator list $Sm[j]$ is searched from $j=0$ to $M-1$. 
For each $Sm[j]\not=0$, a counter $p$ of the number (minus 1) of 
bond-operators encountered is incremented by $1$ and the bond $b=Sm[j]/2$ is 
extracted, giving also the corresponding sites $s_0=Site[1,b]$ and 
$s_1=Site[2,b]$. Links can be set whenever these sites have already been 
encountered, i.e., for $a=0,1$, if $Last(s_a) \not=-1$, $Link[4p+a]=
Last[s_a]$ and $Link[Last[s_a]]=4p+a$. The last occurrence is updated to 
$Last[s_a]=4p+a+2$. If, on the other hand, $Last(s_a)=-1$, only the last 
and first occurrences are recorded, i.e., $Last[s_a]=4p+a+2$ and 
$First[s_a]=4p+a$. The spin list $Spin$ is propagated whenever 
off-diagonal operators are encountered, so that the vertex types $Vtx[p]$ 
can be recorded (using a map from four leg states to the integers 
$1,\ldots,6$). After the whole list $Sm$ has been traversed the list of first 
occurrences is used in order to connect the links across the propagation 
boundary  i.e., for each $s$ for which $Last[s] \not=-1$, $Link[Last[s]]=
First[s]$ and $Link[First[s]]=Last[s]$. 

The loop update is repeated $N_l$ times. Each loop starts at a random
position $j_0 \in \lbrace 0,\ldots,4n-1\rbrace$ in the list $Link$.
We will move in $Link$ and the current position will be referred
to as $j$. We hence begin at $j=j_0$ and keep $j_0$ in order to check at 
each stage whether the loop has closed or not. The current position 
corresponds to vertex number $p=j/4$ and the leg index is $l_i=MOD(j,4)$ 
(we can now for convenience number the legs $0,\ldots,3$). This 
is the entrance leg, and the vertex type is $Vtx[p]$.  The exit probabilities  
given the entrance leg depend on the vertex type and should be stored in a 
pre-generated table. It is convenient to use a list of cumulative exit 
probabilities instead of the individual probabilities, so that for a given 
entrance leg $l_i$ the exit leg can be obtained by successively comparing the 
cumulative probabilities $Prob[l_e,l_i,Vtx[p]]$ for exiting at leg 
$l_e=0,\ldots,3$ with a random number in the range $[0,1]$. A corresponding 
list with updated vertex types is also stored, so that after the exit leg 
has been fixed the vertex is updated as $Vtx(p) \to NewVtx[l_e,l_i,Vtx[p]]$. 
After this, the current position in $Link$ is changed to the one corresponding 
to the exit leg, i.e., $j \to 4p+l_e$. The loop closes at this stage if 
$j=j_0$. If it does not close, we move to the leg linked to $j$, i.e., 
$j \to Link[j]$. The loop closes also at this stage if $j=j_0$. The two 
different types of closings, from within the same vertex or from a different 
vertex, are illustrated in Fig.~\ref{closings}.

After all the $N_l$ loops have been constructed this way, the updated
vertex list $Vtx$ is mapped onto the corresponding new operator list
$Sm$. The bond-indices do not change, and therefore one can simply cycle 
through the positions $j=0,\ldots,M-1$ in the old list one-by-one and for each 
non-zero occurrence extract the bond $b=Sm[j]/2$ and increment an operator 
counter $p \to p+1$ (the corresponding position in the vertex list $Vtx$). 
The operator-type, diagonal or off-diagonal, can be coded in a list
$OpType[v]=0,1$, where $v=1,\ldots ,6$ is the vertex type and $0,1$ 
correspond to diagonal and off-diagonal, respectively. The updated operator 
element is then $Sm[j]=2b+OpType[Vtx[p]]$. The spin list $Spin$ is updated 
using the list of first occurrences that was generated during the construction
of the linked list. For each site $s$, if $First[s]=-1$ no
operator acts on that site and the spin can be flipped, $Spin[s] \to
-Spin[s]$, with probability $1/2$. Otherwise, the updated spin state
is obtained by extracting the vertex number $p=First[s]/4$ and the leg
$l=MOD(First[s],4)$ corresponding to the site in question. The corresponding 
spin state can be stored as a pre-generated map, so that
$Spin[s] \to LegSpin[l,Vtx[p]]$.

We have now described all the basic procedures involved in carrying out one 
MCS using the general operator-loop update. In the special ``deterministic'' 
cases, where the exit leg is given uniquely by the entrance leg, a number of 
rather self-evident and trivial simplifications are possible (see discussion
in Sec.~II-D).

The possibility of aborting loop updates that become excessively long can 
be simply taken into account by exiting the loop update routine without 
mapping the already accomplished changes in the vertex list $Vtx$ back into 
a new operator list $Sm$ and state $Spin$. For the XXZ-model the loops 
typically do not become excessively long in practice however, as was 
demonstrated in a few examples in Sec.~V. 

The expansion cut-off $M$ is adjusted during equilibration
of the simulation by keeping it at $a\times n_{\rm max}$ where $n_{\rm max}$ 
is the largest $n$ reached so far in the simulation and a suitable value for 
the factor is $a \approx 1.25$. The number of loops $N_l$ is also adjusted 
during equilibration, to keep the average total number of vertices visited
in one MCS close to some reasonable number, e.g., $2\langle n\rangle$, as 
discussed in Sec.~II B.  We will not discuss the procedures 
for measuring operator expectation values here, but published forms for 
several types of estimators \cite{sse2,sse4,dorneich2} can be easily 
translated into the data structures used above.

\end{document}